\documentclass[journal, 12pt, draftclsnofoot, onecolumn]{IEEEtran}
\IEEEoverridecommandlockouts

\usepackage{amsfonts,amssymb}
\usepackage{ifpdf}
\usepackage{cite}
\usepackage{stfloats}
\usepackage{bm}
\usepackage{color,xcolor}
\usepackage{subfigure}

\usepackage{algorithm,algpseudocode}

\usepackage[outdir=./]{epstopdf}

\ifCLASSINFOpdf
\usepackage[pdftex]{graphicx}
\graphicspath{{../pdf/}{../jpeg/}}
\DeclareGraphicsExtensions{.pdf,.jpeg,.png}
\else
\usepackage[dvips]{graphicx}

\graphicspath{{../eps/}}

\DeclareGraphicsExtensions{.eps}
\fi

\usepackage[cmex10]{amsmath}

\usepackage{makecell}

\allowdisplaybreaks [4]
\begin{document}
	
	\title{Intelligent Reflecting Surface Empowered Self-Interference Cancellation in Full-Duplex Systems}
	\author{Chi~Qiu, Meng~Hua, Qingqing~Wu, Wen~Chen, Shaodan~Ma, Fen~Hou, Derrick~Wing~Kwan~Ng,~\IEEEmembership{Fellow,~IEEE}, and A.~Lee Swindlehurst,~\IEEEmembership{Fellow,~IEEE}
		\thanks{
			C. Qiu, M. Hua, S. Ma and F. Hou are with the State Key Laboratory of Internet of Things for Smart City, University of Macau, Macau, 999078, China (e-mail: yc17434@connect.um.edu.mo; menghua@um.edu.mo; shaodanma@um.edu.mo; fenhou@um.edu.mo).	Q. Wu and W. Chen are with the Department of Electronic Engineering, Shanghai Institute of Advanced Communications and Data Sciences, Shanghai Jiao Tong University, Minhang 200240, China (e-mail: qingqingwu@sjtu.edu.cn; wenchen@sjtu.edu.cn). D. W. K. Ng is with the School of Electrical Engineering and
			Telecommunications, University of New South Wales, Sydney, Australia (e-mail: w.k.ng@unsw.edu.au).	L. Swindlehurst is with the Center for Pervasive Communications and Computing, Henry Samueli School of Engineering, University of California, Irvine, CA 92697 USA (e-mail: swindle@uci.edu)
	}}%
	\maketitle
	\vspace{-1.5cm}
	\begin{abstract}
		Compared with traditional half-duplex wireless systems, the application of emerging full-duplex (FD) technology can potentially double the system capacity theoretically. However, conventional techniques for suppressing self-interference (SI) adopted in FD systems require exceedingly high power consumption and expensive hardware. In this paper, we consider employing an intelligent reflecting surface (IRS) in the proximity of an FD base station (BS) to mitigate SI for simultaneously receiving data from uplink users and transmitting information to downlink users. The objective considered is to maximize the weighted sum-rate of the system by jointly optimizing the IRS phase shifts, the BS transmit beamformers, and the transmit power of the uplink users. To visualize the role of the IRS in SI cancellation by isolating other interference, we first study a simple scenario with one downlink user and one uplink user. To address the formulated non-convex problem, a low-complexity algorithm based on successive convex
		approximation is proposed. For the more general case considering multiple downlink and uplink users, an efficient alternating optimization algorithm based on element-wise optimization is proposed. Numerical results demonstrate that the FD system with the proposed schemes can achieve a larger gain over the half-duplex system, and the IRS is able to achieve a balance between suppressing SI and providing beamforming gain. 
		
	\end{abstract}

	\section{Introduction}

With high potential to increase system capacity, full duplex (FD) transmissions have recently attracted significant interest. Unlike the half-duplex (HD) counterparts, an FD wireless transceiver can theoretically transmit and receive signals simultaneously to double the system capacity if the inherent self-interference (SI) can be perfectly canceled \cite{RF1}. Although the advantages are predictable, the practical implementation of such FD systems is hindered by numerous difficulties and there are still various technical problems that need to be solved. In particular, the main obstacle to the implementation of practical FD systems is the strong SI that occurs when the transmit and receive antennas interact one another. More specifically, the received signals from uplink (UL) users are overwhelmed by the downlink (DL) signals. Indeed, the amount of SI suppression at the transceivers, which is typically constrained in practice, determines how well FD systems operate. For example, it was shown in \cite{spectral_efficiency} that the performance of an FD system might be poorer than that of the HD counterpart in the estimate of large residual SI. In recent years, several advancements have been reported in the literature on effective hardware design for SI interference suppression. Specifically, existing SI mitigation methods mainly rely on deploying additional active radio frequency (RF) chains \cite{RF1,RF2,RF3,RF4}, and typically involve several steps. First, an RF circulator is utilized to direct the transmitted signal to the antenna and the received signal to the receiver chain. Analog cancellation is then performed by estimating and subtracting the leaked self-interference from the received signal. Next, digital cancellation is employed to further mitigate the remaining self-interference by utilizing adaptive filtering and cancellation algorithms. These algorithms estimate and subtract the self-interference based on the received signal characteristics and the cancellation obtained from the analog domain. In \cite{RF4}, it  was proposed to use an auxiliary transmitter as  a novel nonlinear
SI cancellation approach, which
effectively solves the problems of in-phase (I) and
quadrature (Q) imbalance and nonlinear
distortions. Since then, studies on FD-enabled wireless communication have been carried out for different applications, e.g., \cite{FD1, FD2,FD3}. Despite numerous efforts that have been devoted, such SI mitigation techniques entail both high hardware costs and power consumption.

Instead of applying the RF-chain-based technique, an intelligent reflecting surface (IRS) offers a novel option for effective SI mitigation. IRS technology has recently shown promise in improving the efficiency of wireless communication systems. An IRS is a meta-surface made up of a large amount of passive reflecting elements, each of which can alter the phases and amplitudes of the incoming signals independently. As such, an IRS can be programmed to flexibly and adaptively reflect the incoming signals in a desired manner. For instance, the signals reflected by the IRS can be superimposed constructively at an intended receiver or destructively at an undesired receiver with that from the direct path via passive beamforming. In addition, the advantages of low-dimensional, flexible, and compact allow the IRS to be seamlessly integrated into existing wireless systems \cite{GR,IRS1,IRS2}. In particular, both the low hardware cost and the passive nature of an IRS facilitate sustainable development of practical wireless networks and thus it has been exploited for various applications, such as coverage extension \cite{coverage_extension1,coverage_extension2,coverage_extension3}, mobile-edge computing \cite{MEC1,MEC2,MEC3}, integrated sensing and communication systems \cite{ISAC1,ISAC2,ISAC3}, multi-cell systems \cite{multi_cell1,multi_cell2,multi_cell3}, etc. However, the above works have primarily focused on the capability of an IRS to either increase the received signal strength or mitigate the intra- /inter-cell interference at the receiver \cite{interference_cancellation}. In fact, an IRS is also appealing for SI cancellation by its tunable nature. The signal path from the transmit antennas to the receive antennas, i.e., the loop-interference (LI) link, within an FD node can be treated as a wireless communication channel. As such, by properly tuning the IRS phase shifts, the interference induced by the leakage of the transmitted signal can be potentially suppressed at the receiver.

Recently, several works have been published investigating IRS-assisted SI cancellation in FD systems to address the SI problems in a cost-effective manner \cite{FD_SI7,FD_SI3_omni,FD_SI6,FD_SI2_NF,FD_SI5,FD_SI4_UL,FD_SI1_IB,FD_SI8_connectivity}, reducing both the hardware cost and computational complexity compared with standard RF-chain-based SI cancellation. For instance, in \cite{FD_SI7}, an FD testbed involving a $256$-element IRS prototype was shown to achieve a $59$ dB reduction in SI. Also, various system setups have been investigated for joint IRS reflection coefficients and BS beamforming design to facilitate FD communications. For example, for point-to-point communication system with one FD transceiver communicating with one user, joint design algorithms were proposed in \cite{FD_SI3_omni} for two types of intelligent omni-surfaces to maximize the data throughput and to minimize the SI power, respectively. The model was then extended in \cite{FD_SI6} to include multiple FD links communicating simultaneously with the help of an IRS via adaptive SI and inter-node interference cancellation. Furthermore, the authors in \cite{FD_SI2_NF} introduced two cooperative near-field IRSs into the point-to-point communication system for the weighted sum-rate (WSR) maximization. In addition, the bit error probability for a broadcast system with one DL user and one UL user was analyzed theoretically  in \cite{FD_SI5}. Besides, the analysis was extended to the system with multiple DL and UL users in \cite{FD_SI4_UL,FD_SI1_IB,FD_SI8_connectivity} for different performance metrics. However, the existing literature lacks a comprehensive model that considers both DL and UL communication, as well as the mutual interference between them,  leaving a gap in the understanding of the FD systems. Furthermore, it remains unexplored how to design the IRS phase shifts to strike a balance between SI cancellation and information transmission enhancement in such systems. To facilitate the practical implementation of the IRS-empowered SI cancellation in FD systems, it becomes imperative to establish a foundational framework with a low-complexity design, which is still missing in the literature.

Motivated by the above, in this paper we explore the potential gains of an IRS-assisted FD system in which a base station (BS), with the help of an IRS, receives signals from a set of UL users and simultaneously transmits signals to a set of DL users. The design problem is further complicated by the coupled multi-user interference (MUI) and co-channel interference (CCI). For this reason, the IRS is used not only to cancel SI but also to concurrently improve UL and DL channel quality. Our work is significantly different from \cite{FD_SI1_IB} mainly in two aspects. First, CCI from the UL users to the DL transmission was not considered in \cite{FD_SI1_IB}, and including this factor increases the difficulty of the problem. In our work, we consider not only the CCI but also the joint design of the DL/UL transmission. We propose two new resource allocation algorithms catered to the newly formulated problems that balance the role of the IRS between SI cancellation and mitigation of the MUI and CCI. Second, to shed light on the role of IRS in SI cancellation, we study detailed aspects of a simplified setting, where the BS is implemented with a pair of transmit/receive antennas, serving one DL user and one UL user, such that the effect of other interference can be isolated. The proposed algorithm for the single-user scenario has low complexity, making it more computationally efficient and easier to implement in practical systems. The following is a summary of this paper's significant contributions:\begin{itemize}
		\item We investigate an IRS-aided FD wireless communication system whose goal is to maximize the WSR under the power constraints, taking into account the SI, MUI, and CCI. Specifically, we employ an IRS in the vicinity of the FD BS to offer a new option for SI mitigation. The IRS serves not only to cancel the SI, but also to achieve passive beamforming gain for information transmission. In the context of IRS-empowered SI mitigation, it is crucial to investigate the behavior of IRS by isolating the phase-shift design from other interference. We first study a simplified setting, and propose an algorithm with low complexity. Specifically, we propose a low-complexity algorithm based on successive convex approximation (SCA) that recasts the problem as a second-order cone program (SOCP).
		
		\item For the general case with multiple DL and UL users, the formulated problem becomes more challenging due to the presence of MUI. To tackle this difficulty, an element-wise (EW) alternating optimization (AO) is proposed. Specifically, we decompose the original optimization problem into two subproblems: DL/UL transmission and IRS phase shift optimization, respectively. Then, by iteratively optimizing each subproblem, convergence is achieved. Specifically, for the DL/UL transmission subproblem, semidefinite relaxation (SDR) is adopted and the optimality of the relaxed solution is proved.
		
		\item Simulation results demonstrate the benefit of deploying IRS in improving the performance of the considered FD wireless communication system. Specifically, under the simple system setup, we see that the IRS can completely eliminate the SI at the BS. Furthermore, with the help of the IRS, the MUI and CCI can also be mitigated for enhancing the transmission quality. It is demonstrated that the proposed scheme in the FD system outperforms its traditional HD counterpart and other baseline schemes, even in the presence of strong SI. Moreover, the IRS shows its superiority in balancing the SI suppression and transmission enhancement.
	\end{itemize}
	
	The remainder of this paper is organized as follows. In Section II, the system model is presented and the WSR maximization problem is formulated for the considered IRS-assisted FD system. Section III studies a simple case with one DL and one UL user and proposes a low-complexity SCA-based algorithm. In Section IV, we propose an EW-based AO algorithm for the general case. Section V provides the numerical results to validate the effectiveness of the proposed algorithms and finally, Section VI brings the conclusion.
	
	\emph{Notations}: Throughout the paper, we denote matrices and vectors by boldface upper-case and lower-case letters, respectively. Let $\mathbb{C}^{ a \times b}$ denote complex matrices of dimensions $a\times b$. For a complex-valued
	vector $\bm x$,  $(\bm x)^H$ and $\operatorname{diag}(\bm x)$  represent the Hermitian
	transpose and the diagonal matrix formed by the elements of $\mathbf{x}$, respectively. For a matrix $\bm X$,  let $\operatorname{tr}(\bm X)$, $\operatorname{rank}(\bm X)$, and $\operatorname{vec}(\bm X)$ denote the trace, the rank, and the vector formed by stacking the columns of $\bm X$, respectively. In addtion, $\boldsymbol{X} \succeq \bm 0$ implies that $\bm X$ is positive semidefinite, where $\mathbf{0}$ represents the matrix of all zeros. $\nabla_{\mathbf{X}} f(\mathbf{X})$ denotes the gradient of $f\left(\mathbf{X}\right)$. For a complex scalar $x$, $|x|$ and $\operatorname{Re}\{x\}$ denote the absolute value and the real part of $x$, respectively.   $\otimes$ represents the Kronecker product. For a circularly symmetric complex Gaussian distributed random variable $x$ with mean $\mu$ and variance $\sigma^2$, it is denoted as $x \sim \mathcal{C} \mathcal{N}\left(\mu, \sigma^2\right)$. $\mathcal{O}(\cdot)$ denotes the big-O computational complexity.
	
	\section{System Model and Problem Formulation}
	\subsection{System Model}
	\begin{figure}[!t]
		\centerline{\includegraphics[width=5.3in]{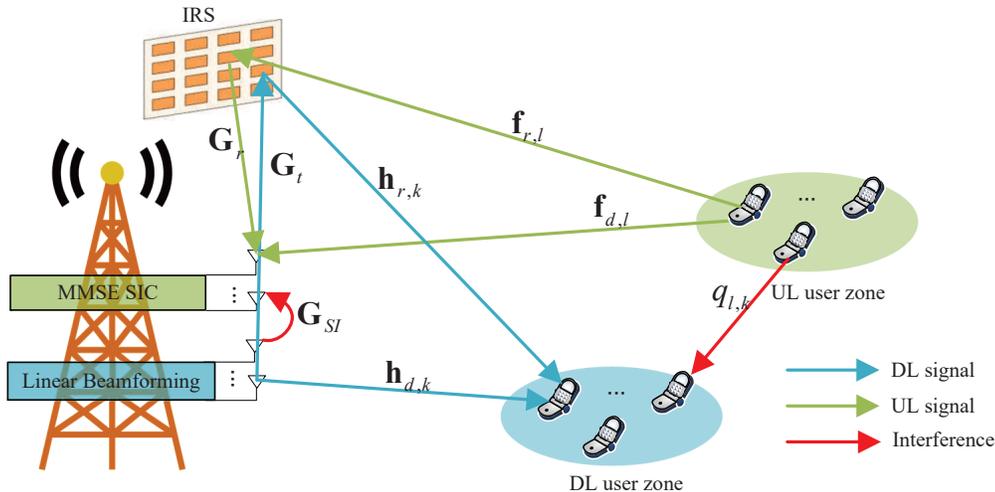}}
		\vspace{-0.5cm}
		\caption{An IRS-aided FD system for serving $K$ DL users and $L$ UL users.} 
		\label{system_model}
		\vspace{-1cm}
	\end{figure}
	
	As illustrated in Fig. \ref{system_model}, we consider an FD wireless communication system, which includes a BS, $K$ DL users, $L$ UL users, and an IRS. Specifically, we denote the sets of DL and UL users by $\mathcal{K}=\{1,\dots,K\}$ and $\mathcal{L}=\{1,\dots,L\}$, respectively. We assume that the 
	BS is equipped with a total number of $N_{\text{T}}+N_{\text{R}}$ antennas operating in FD mode. The BS broadcasts the signals to the DL users with $N_{\text{T}}$ transmit antennas and simultaneously receives the signals from the UL users with $N_{\text{R}}$ receive antennas over the same frequency band. We assume that the users are all with single-antenna. For the IRS, it is implemented with $M$ passive reflecting elements, being deployed in the vicinity of the BS.
	In contrast to user-side IRS, BS-side IRS offers several advantages. First, the BS-side IRS serves both DL and UL user groups, thus significantly expanding the IRS's reflected signal coverage. Second, by strategically placing the IRS near the BS, the power of the reflected SI link is comparable in strength to the LI channel, and thus the potential SI in the FD system is more likely to be canceled\footnote{
		When the intelligent reflecting surface (IRS) is not in the vicinity of the BS, the transmission power for the DL may become high, leading to a strong SI that can significantly degrade the performance of the UL user group, particularly if the latter is closer to the BS than the DL one. In contrast, if the DL user group is closer to the BS than the UL user group, the IRS will have a greater impact on the DL transmission, resulting in zero UL rate. Therefore, for achieving satisfactory performance, it is preferable to deploy the IRS in proximity to the BS.}.  Denote the IRS reflection-coefficient matrix by $\mathbf{\Phi} = \operatorname{diag}\left(e^{j\theta_{1}},\dots,e^{j\theta_{M}}\right)\in \mathbb{C}^{M\times M}$, where  the reflecting amplitude is set to be 1 and $\theta_{m} \in [0,2\pi), m\in \mathcal{M}=\{1,\dots,M\}$ represents the phase shift at the $m$-th IRS reflecting element. 
	
To study the system performance uppper bound, we assume that perfect channel state information (CSI) of all channels at both the BS and users \cite{CSI}. With assumed quasi-static flat-fading environment, for the DL channels, we denote the baseband equivalent channels from the BS's transmit antennas to the IRS, from the IRS to DL user $k$, $k\in \mathcal{K}$, and from the BS's transmit antennas to DL user $k$ by $\mathbf{G}_t\in\mathbb{C}^{M\times N_{\text{T}}}$, $\mathbf{h}_{r,k}^H\in\mathbb{C}^{1\times M}$, and $\mathbf{h}_{d,k}^H\in\mathbb{C}^{1\times N_{\text{T}}}$, respectively. For the UL channels, we denote the baseband equivalent channels from the IRS to the BS's receive antennas, from  UL user $l$, $l\in\mathcal{L}$ to the IRS, and from UL user $l$ to the BS's receive antennas by $\mathbf{G}_r\in\mathbb{C}^{N_{\text{R}}\times M}$, $\mathbf{f}_{r,l}\in\mathbb{C}^{M\times 1}$, and $\mathbf{f}_{d,l}\in\mathbb{C}^{N_{\text{R}}\times 1}$, respectively. In addition, let $\mathbf{G}_{\text{SI}} \in \mathbb{C}^{N_{\text{R}
		}\times N_{\text{T}}}$ denote the LI channel from the transmit antennas to the receive antennas at the BS.

Let $s_{\text{D},k}$ denote the transmitted symbol for DL user $k$, which satisfies $\mathbb{E}\left\{\left|s_{\text{D},k}\right|^2\right\}=1$. Since we adopt the conventional linear beamforming for DL user $k$, the information-carrying symbol $s_{\text{D},k}$ is multiplied by the beamforming vector $\mathbf{w}_k \in \mathbb{C}^{N_{\text{T}}\times 1}$. Accordingly, the received signal at DL user $k$ is given by
	\begin{equation}
			y_{\text{D},k}=\left(\mathbf{h}_{d,k}^H+\mathbf{h}_{r,k}^H\boldsymbol{\Theta}\mathbf{G}_{t}\right)\mathbf{w}_ks_{\text{D},k}+\underbrace{\sum_{i\neq k}^{K}\left(\mathbf{h}_{d,k}^H+\mathbf{h}_{r,k}^H\boldsymbol{\Theta}\mathbf{G}_{t}\right)\mathbf{w}_is_{\text{D},i}}_{\text{MUI}}+\underbrace{\sum_{l=1}^{L}q_{l,k}s_{\text{U},l}+n_{\text{D},k}}_{\text{CCI}},
	\end{equation}where $s_{\text{U},l}$ is the symbol transmitted by UL user $l$ and $n_{\text{D},k} \sim \mathcal{C} \mathcal{N}\left(0, \sigma_n^2\right)$ is additive white Gaussian noise (AWGN) with $\sigma_n^2$ being the noise variance. Since the IRS is sufficiently far from the users, we can infer that the average signal power through the ``UL user - IRS - DL user" is negligible and thus ignored. The signal-to-interference-plus-noise ratio (SINR) of DL user $k$ is expressed as
	\begin{equation}
			\gamma_{\text{D},k}=\frac{|\bar{\mathbf{h}}_{k}^H\mathbf{w}_{k}|^2}{\sum_{i\neq k}^{K}|\bar{\mathbf{h}}_{k}^H\mathbf{w}_i|^2+\sum_{l=1}^{L}p_{U,l}|q_{l,k}|^2+\sigma_n^2},
	\end{equation}
	where $\bar{\mathbf{h}}_{k}^H=\mathbf{h}_{d,k}^H+\mathbf{h}_{r,k}^H\boldsymbol{\Theta}\mathbf{G}_{t}$ and $\mathbb{E}\left\{\left|s_{\text{U},l}\right|^2\right\}=p_{\text{U},l}$, $l=1,\dots,L$, is the transmit power of UL user $l$. Thus, the sum-rate of  DL transmission is given by 
		\begin{equation}
			R_{\text{D}}=\sum_{k=1}^{K}\operatorname{log}_2\left(1+\gamma_{\text{D},k}\right).
		\end{equation}
	
	On the other hand, for the UL transmission, the received signal at the BS's receive antennas is given by
	\begin{equation}
		\mathbf{y}_{\text{U}}=\sum_{l=1}^{L}\left(\mathbf{f}_{d,l}+\mathbf{G}_{r}\boldsymbol{\Theta}\mathbf{f}_{r,l}\right)s_{\text{U},l}+\underbrace{\sum_{k=1}^{K}\left(\mathbf{G}_{\text{SI}}+\mathbf{G}_{r}\boldsymbol{\Theta}\mathbf{G}_{t}\right)\mathbf{w}_ks_{\text{D},k}}_{\text{SI}}+\mathbf{n}_{\text{U}},
	\end{equation} where $\mathbf{n}_{\text{U}} \sim \mathcal{CN}\left(0, \sigma_n^2 \mathbf{I}_{N_{\text{R}}}\right)$ is the received AWGN. In order to achieve the UL capacity, the minimum mean square error and successive interference cancellation (MMSE-SIC) technique is adopted, in which the order of decoding is typically chosen based on the users estimated SINR \cite{MMSE_SIC}. Since the effective channel gain depends on the IRS phase shifts, we adopt a strategy in which the decoding order is determined by the gain of the corresponding direct channel, and the user with the highest direct link gain is decoded first\footnote{Although such a strategy is adopted, the UL sum-rate expression \eqref{6} is indeed irrelevant to the decoding order, thus the proposed algorithms and results are applicable to any given decoding order.}.  By adopting MMSE-SIC for the UL transmission, the SINR of UL user $l$ is given by
	\begin{equation}
		\gamma_{\text{U},l}=p_{\text{U},l}\bar{\mathbf{f}}_{l}^H\left(\sum_{i>l}^{L}p_{\text{U},i}\bar{\mathbf{f}}_{i}\bar{\mathbf{f}}_{i}^H+\sum_{k=1}^{K}\bar{\boldsymbol{G}}\mathbf{w}_k\mathbf{w}_k^H\bar{\boldsymbol{G}}^H+\sigma_n^2\mathbf{I}\right)^{-1}\bar{\mathbf{f}}_{l},
	\end{equation}
	where $\bar{\mathbf{f}}_{l}=\mathbf{f}_{d,l}+\mathbf{G}_{r}\boldsymbol{\Theta}\mathbf{f}_{r,l}$ and $\bar{\boldsymbol{G}}=\mathbf{G}_{\text{SI}}+\mathbf{G}_{r}\boldsymbol{\Theta}\mathbf{G}_{t}$. Consequently, the achievable sum-rate of UL transmission rate is given by
		\begin{align}\label{6}
			R_{\text{U}}&=\sum_{l=1}^{L}\operatorname{log}_2\left(1+\gamma_{\text{U},l}\right)\\
			&=\operatorname{log}_2\frac{|\sum_{l=1}^{L}p_{U,l}\bar{\mathbf{f}}_{l}\bar{\mathbf{f}}_{l}^{H}+\sum_{k=1}^{K}\bar{\boldsymbol{G}}\mathbf{w}_k\mathbf{w}_k^H\bar{\boldsymbol{G}}^H+\sigma^2_n\mathbf{I}|}{|\sum_{k=1}^{K}\bar{\boldsymbol{G}}\mathbf{w}_k\mathbf{w}_k^H\bar{\boldsymbol{G}}^H+\sigma^2_n\mathbf{I}|}.
		\end{align}

	\subsection{Problem Formulation}
	We see from the above that the MUI, CCI, and SI have significant impacts on the system throughput. Thus, in this paper, we aim to maximize the WSR of the system by a joint design of the DL and UL transmission parameters and the IRS passive beamformers, taking into account BS and per-user power constraints. Define $\mathbf{w}=\{\mathbf{w}_k,k\in\mathcal{K}\}$, $p=\{p_{\text{U},l}, l\in\mathcal{L}\}$, and $\mathbf{v}=\left[e^{j\theta_1},\dots,e^{j\theta_M}\right]^T$. The WSR problem is formulated as  \vspace{-0.3cm}
	\begin{subequations}\label{problem1}
			\begin{align}
				&\max_{\mathbf{v},\mathbf{w},p} ~
				\alpha_{\text{D}}R_{\text{D}}+\alpha_{\text{U}}R_{\text{U}}\label{7a}\\
				&~\text{s.t.} ~~ 0\leq p_{\text{U},l}\leq p_{\text{U},l}^{\text{max}},l\in\mathcal{L}\label{p_constraint},\\
				&~\qquad\sum_{k=1}^{K}\operatorname{Tr}\left(\mathbf{w}_k\mathbf{w}_k^H\right)\leq P_{\text{BS}}^{\text{max}},\label{constraint2}\\
				&~\qquad|[\mathbf{v}]_m|=1, m \in \mathcal{M}\label{constraint3},
			\end{align}
	\end{subequations}where $\alpha_{\text{D}}\geq0$ and $\alpha_{\text{U}}\geq0$ denote weighting factors that balance the importance of the DL and UL transmission, $p_{\text{U},l}^{\text{max}}$ in \eqref{p_constraint} denotes the power budget at UL user $l$, and $P_{\text{BS}}^{\text{max}}$ in \eqref{constraint2} denotes the power budget at the BS. Constraint  \eqref{constraint3} ensures the unit-modulus phase shift at the IRS. Problem \eqref{problem1} is non-convex and challenging to solve due to the coupling of variables in the objective function \eqref{7a}. To shed light on the IRS-empowered SI mitigation,  we study the case with only one DL user and one UL user as a simplified yet fundamental setting in the next section. Subsequently, an SCA-based algorithm is proposed to solve the simplified problem,  which provides a useful and low-complexity framework  in the context of practical implementation.  In Section IV, we extend the solution to the general case involving BS transmit beamforming design.

	\section{A Simple System Setup with a DL user and a UL user}
	
	In this section, we study the simple case in which the FD BS is equipped with one transmit antenna and one receive antenna, i.e., $N_{\text{T}}=N_{\text{R}}=1$, serving one DL user and one UL user, i.e., $K=L=1$.  In this case, no MUI is present in the DL channel, thus the SINRs of the DL user and the UL user are respectively given by (after dropping the user index)
	\begin{align}
		&\gamma_{\text{D}}=\frac{P_{\text{BS}}\left|h_{d}+\mathbf{h}_{r}^H\boldsymbol{\Theta}\mathbf{g}_t\right|^2}{p_\text{U}\left|q\right|^2+\sigma_n^2}=\frac{P_{\text{BS}}\left|h_{d}+\mathbf{h}_{r}^H\boldsymbol{\Theta}\mathbf{g}_t\right|^2}{p_\text{U}\left|q\right|^2+\sigma_n^2},\\
		&\gamma_{\text{U}}=\frac{p_{\text{U}}\left|f_{d}+\mathbf{g}_{r}^H\boldsymbol{\Theta}\mathbf{f}_{r}\right|^2}{P_{\text{BS}}\left|g_{\text{SI}}+\mathbf{g}_r^H\boldsymbol{\Theta}\mathbf{g}_t\right|^2+\sigma_n^2},
	\end{align} where $\mathbf{g}_t$ and $\mathbf{g}_r$ denote the baseband equivalent channels between the BS transmit/receive antennas and the IRS, and $g_{\text{SI}}$ denotes the LI channel of the BS. Then, problem \eqref{problem1} is simplified to\begin{subequations}\label{problem2}
		\begin{align}
			&\max_{P_{\text{BS}},p_{\text{U}},\mathbf{v}}~ \alpha_{\text{D}}\operatorname{log}_2\left(1+\gamma_{\text{D}}\right)+\alpha_{\text{U}}\operatorname{log}_2\left(1+\gamma_{\text{U}}\right)\\
			&~\text{s.t.} ~~ P_{\text{BS}}\leq P_{\text{BS}}^{\text{max}},\label{35b}\\
			&~\qquad p_{\text{U}}\leq p_{\text{U}}^{\text{max}},\label{35c}\\
			&~\qquad\eqref{constraint3}.
		\end{align}
	\end{subequations}
	Though problem \eqref{problem2} is much simplified, it is still non-convex. In the following, we detail the algorithm that  solves it suboptimally.
	
	By introducing slack variables $t_{\text{D}}$, $t_{\text{U}}$, $\kappa_{\text{D}}$, and $\kappa_{\text{U}}$, problem \eqref{problem2} can be reformulated as\begin{subequations}\label{problem2_2}
		\begin{align}
			&\max_{P_{\text{BS}},p_{\text{U}},\mathbf{v}}~ \alpha_{\text{D}}\operatorname{log}_2\left(\kappa_{\text{D}}\right)+\alpha_{\text{U}}\operatorname{log}_2\left(\kappa_{\text{U}}\right)\\
			&~\text{s.t.} ~~  t_{\text{D}}^2\leq\frac{P_{\text{BS}}\left|h_{d,k}+\mathbf{h}_{r,k}\boldsymbol{\Theta}\mathbf{g}_t\right|^2}{p_{\text{U}}\left|q\right|^2+\sigma_n^2} \label{24},\\
			&~\qquad t_\text{U}^2\leq\frac{p_{\text{U}}\left|f_{d}+\mathbf{g}_{r}^H\boldsymbol{\Theta}\mathbf{f}_{r}\right|^2}{P_{\text{BS}}\left|g_{\text{SI}}+\mathbf{g}_r^H\boldsymbol{\Theta}\mathbf{g}_t\right|^2+\sigma_n^2}\label{25},\\
			&~\qquad\kappa_{\text{D}}\leq 1+t_{\text{D}}^2, \kappa_{\text{U}}\leq 1+t_{\text{U}}^2, \label{24d}\\
			&~\qquad\eqref{constraint3},\eqref{35b},\eqref{35c}.
		\end{align}
	\end{subequations}
	Note that the inequality constraints \eqref{24}-\eqref{24d} are always active at the optimal solution, which can be proved by contradiction and is omitted here for brevity. Problem \eqref{problem2_2} is thus equivalent to problem \eqref{problem2}. Let $|h_{d,k}+\mathbf{h}_{r,k}^H\boldsymbol{\Theta}\mathbf{g}_t|^2=|\hat{\mathbf{h}}^H\hat{\mathbf{v}}|^2$, where $\hat{\mathbf{h}}^H=[\mathbf{h}_{r,k}^H\text{diag}\left(\mathbf{g}_t\right) \ h_{d,k}]$ and $\hat{\mathbf{v}}=[\mathbf{v}^H \ 1]^H$. Define $\hat{\mathbf{H}}=\hat{\mathbf{h}}\hat{\mathbf{h}}^H$. Then, constraint \eqref{24} can be expressed as\begin{equation}
		p_{\text{U}}\left|q\right|^2+\sigma^2\leq \frac{P_{\text{BS}}\hat{\mathbf{v}}^H\hat{\mathbf{H}}\hat{\mathbf{v}}}{t_{\text{D}}^2}.\label{32a}
	\end{equation}Then, by further introducing slack variable $\beta_{\text{D}}$, we have\begin{equation}
	t_{\text{D}}^2\leq\beta_{\text{D}}P_{\text{BS}}\label{42},
\end{equation}such that we can rewrite constraint \eqref{24} into	\begin{equation}
p_{\text{U}}\left|q\right|^2+\sigma^2\leq \frac{\hat{\mathbf{v}}^H\hat{\mathbf{H}}\hat{\mathbf{v}}}{\beta_{\text{D}}}.\label{28}
\end{equation}The key observation is that the right-hand-side (RHS) of \eqref{28} is jointly convex with respect to (w.r.t.) $\hat{\mathbf{v}}$ and $\beta_{\text{D}}$. Motivated by the fact that any convex function is lower bounded by its first-order Taylor expansion, we apply the SCA technique to tackle it. Specifically, with given feasible points $\hat{\mathbf{v}}^{r}$ and $\beta^r_{\text{D}}$, we derive the following lower bound\begin{equation}
\frac{\hat{\mathbf{v}}^H\hat{\mathbf{H}}\hat{\mathbf{v}}}{\beta_{\text{D}}}\geq \frac{ \operatorname{Re}\left\{\hat{\mathbf{v}}^{H} \mathrm{H} \hat{\mathbf{v}}^r\right\}}{\beta_{\text{D}}^r}- \frac{\hat{\mathbf{v}}^{r,H} \mathrm{H} \hat{\mathbf{v}}^r}{\beta_{\text{D}}^r} -\frac{\hat{\mathbf{v}}^{r,H} \mathrm{H} \hat{\mathbf{v}}^r}{\left(\beta_{\text{U}}^{r}\right)^2} \left(\beta_{\text{D}}-\beta_{\text{D}}^r\right)\triangleq\gamma^{\text{lb}}\left(\hat{\mathbf{v}},\beta_{\text{D}}\right).
\end{equation} Similarly, by further introducing slack variables $\beta_{\text{BS}}$ and $\beta_{\text{U}}$, we have \vspace{-0.2cm}
\begin{equation}
\beta_{\text{BS}}P_{\text{BS}}\leq 1\label{31a},
\end{equation}
\begin{equation}
t_{\text{U}}^2\leq\beta_{\text{U}}p_{\text{U}},\label{18}
\end{equation} and constraints \eqref{25} can be rewritten as\begin{equation}
\frac{\hat{\mathbf{v}}^H\mathbf{G}\hat{\mathbf{v}}}{\beta_{\text{BS}}}+\sigma^2 \leq \frac{\hat{\mathbf{v}}^H\mathbf{F}\hat{\mathbf{v}}}{\beta_{\text{U}}}\label{28a},\\
\end{equation}where $\mathbf{G}=\hat{\mathbf{g}}\hat{\mathbf{g}}^H$, $\mathbf{F}=\hat{\mathbf{f}}\hat{\mathbf{f}}^H$,  $\hat{\mathbf{g}}^H=[\mathbf{g}_{r}^H\text{diag}\left(\mathbf{g}_t\right) \ g_{\text{SI}}]$, and $\hat{\mathbf{f}}^H=[\mathbf{g}_{r}^H\text{diag}\left(\mathbf{f}_r\right) \ f_{d}]$. Again, with given initial points $\hat{\mathbf{v}}^r$
and $\beta_{\text{U}}^r$, the lower bound of $\hat{\mathbf{v}}^H\mathbf{F}\hat{\mathbf{v}}/\beta_{\text{U}}$ can be obtained as\begin{equation}
	\frac{\hat{\mathbf{v}}^H\mathbf{F}\hat{\mathbf{v}}}{\beta_{\text{U}}}\geq  2 \frac{ \operatorname{Re}\left\{\hat{\mathbf{v}}^{H} \mathbf{F} \hat{\mathbf{v}}^r\right\}}{\beta_{\text{U}}^r}- \frac{\hat{\mathbf{v}}^{r,H} \mathbf{F} \hat{\mathbf{v}}^r}{\beta_{\text{U}}^r} -\frac{\hat{\mathbf{v}}^{r,H} \mathbf{F} \hat{\mathbf{v}}^r}{\left(\beta_{\text{U}}^{r}\right)^2} \left(\beta_{\text{U}}-\beta_{\text{U}}^r\right)\triangleq \rho^{\text{lb}}\left(\hat{\mathbf{v}},\beta_{\text{U}}\right).
\end{equation} In addition, from \cite{spectral_efficiency}, the convex upper bound estimate of left-hand-side (LHS) of constraint \eqref{31a} can be found as \begin{equation}
\beta_{\text{BS}}P_{\text{BS}}\leq\frac{1}{2 \psi_{\text{BS}}^{r}} \beta_{\text{BS}}^2+\frac{\psi_{\text{BS}}^{r}}{2} P_{\text{BS}}^2
\triangleq \lambda^{\text{ub}}\left(\beta_{\text{BS}},P_{\text{BS}},\psi_{\text{BS}}^{r}\right),
\end{equation}which holds for every $\psi_{\text{BS}}^r\geq0$. It is straightforward that equality holds when $\psi_{\text{BS}}^{r}=\beta_{\text{BS}}/P_{\text{BS}}$. Finally, for the constraints in \eqref{24d}, we can similarly obtain the corresponding lower bounds with given initial points $t_{\text{D}}^r$ and $t_{\text{U}}^r $, yielding
\vspace{-0.2cm}
		\begin{equation}
			1+t_{\text{D}}^2\geq\eta^{\text{lb}}\left(t_{\text{D}}\right), 1+t_{\text{U}}^2\geq\eta^{\text{lb}}\left(t_{\text{U}}\right),
		\end{equation}	where $\eta^{\text{lb}}\left(x\right)\triangleq1+\left(x^{r}\right)^2+2x^r\left(x-x^r\right)$.

	Therefore, by recasting the hyperbolic constraints \eqref{42} and \eqref{18} based on \cite{socp} and relaxing the non-convex unit-modulus constraint, problem \eqref{problem2} can be approximated as
	\begin{subequations}\label{prob2}
		\begin{align}
			&\max_{\substack{\kappa_{\text{D}},\kappa_{\text{U}},t_{\text{D}},t_{\text{U}},\beta_{\text{D}},\\\beta_{\text{U}},\beta_{\text{BS}},\mathbf{v},P_{\text{BS}},p_{\text{U}}}}~ \alpha_{\text{D}}\operatorname{log}_2\left(\kappa_{\text{D}}\right)+\alpha_{\text{U}}\operatorname{log}_2\left(\kappa_{\text{U}}\right)\\
			&~\text{s.t.}~~ \label{relaxed_constraint3}
			|[\mathbf{v}]_m|\leq1, m \in \mathcal{M}, \\
			&~\qquad p_u\left|q\right|^2+\sigma^2\leq \gamma^{\text{lb}}\left(\hat{\mathbf{v}},\beta_{\text{D}}\right),\\ &\qquad\frac{\hat{\mathbf{v}}^H\mathbf{G}\hat{\mathbf{v}}}{\beta_{\text{BS}}}+\sigma^2\leq \rho^{\text{lb}}\left(\hat{\mathbf{v}},\beta_{\text{U}}\right),\\
			&~\qquad \lambda^{\text{ub}}\left(\beta_{\text{BS}},P_{\text{BS}},\psi_{\text{BS}}^{r}\right)\leq 1,\\
			&~\qquad\kappa_{\text{D}}\leq\eta^{\text{lb}}\left(t_{\text{D}}\right), \kappa_{\text{U}}\leq\eta^{\text{lb}}\left(t_{\text{U}}\right),\\
			&~\qquad \label{46}
			\left\|\left[\begin{array}{c}
				2t_{\text{D}} \\
				\beta_{\text{D}}-P_{\text{BS}}
			\end{array}\right]\right\| \leqslant \beta_{\text{D}}+P_{\text{BS}},\\
			&~\qquad \label{47}
			\left\|\left[\begin{array}{c}
				2t_\text{U} \\
				\beta_{\text{U}}-p_{\text{U}}
			\end{array}\right]\right\| \leqslant \beta_{\text{U}}+p_{\text{U}},\\
			&~\qquad\eqref{35b},\eqref{35c},
		\end{align}
	\end{subequations}
	which is an SOCP and can be solved by existing solvers such as CVX \cite{cvx}. 
	
	\begin{algorithm}[t]
		\caption{SCA-based Algorithm for Solving Problem \eqref{problem2}}\label{algorithm2}
		\begin{algorithmic}[1]
			\State \textbf{Initialize} the IRS phase-shift vector $\mathbf{v}^r$, DL BS power $P_{\text{BS}}^{r}$, UL user power $p_{\text{U}}^r$, $\psi_{\text{BS}}^{r}$, $t_{\text{D}}^{r}$, $t_{\text{U}}^{r}$, $\beta_{\text{D}}^{r}$, $\beta_{\text{U}}^{r}$, $\beta_{\text{BS}}^{r}$, iteration index $r=0$, and threshold $\epsilon$.
			\Repeat
			\State Solve problem \eqref{prob2} to find optimal $\kappa_{\text{D}}^*$, $\kappa_{\text{U}}^*$, $t_{\text{D}}^*$, $t_{\text{U}}^*$, $\beta_{\text{D}}^*$, $\beta_{\text{U}}^*$, $\beta_{\text{BS}}^*$, $\mathbf{v}^*$, $P_{\text{BS}}^*$, $p_{\text{U}}^*$.
			\State Set: $r:=r+1$.
			\State Update: $\psi_{\text{BS}}^{r}:=\beta_{\text{BS}}^*/P_{\text{BS}}^*$, $t_{\text{D}}^r:=t_{\text{D}}^*$, $t_{\text{U}}^r:=t_{\text{U}}^*$, $\beta_{\text{D}}^r:=\beta_{\text{D}}^*$, $\beta_{\text{U}}^r:=\beta_{\text{U}}^*$, $\beta_{\text{BS}}^r:=\beta_{\text{BS}}^*$, $\mathbf{v}^r:=\mathbf{v}^*$, $P_{\text{BS}}^r:=P_{\text{BS}}^*$.
			
			\Until the increase of the objective value of problem \eqref{prob2} is below $\epsilon$.
			\State Reconstruct the IRS phase shifts based on \eqref{unit_modulus}.
			\State Update $ \{\kappa_{\text{D}},\kappa_{\text{U}},t_{\text{D}},t_{\text{U}},\beta_{\text{D}},\beta_{\text{U}},\beta_{\text{BS}},P_{\text{BS}},p_{\text{U}}\}$ based on the reconstructed IRS phase-shift vector.
		\end{algorithmic}
	\end{algorithm}

The overall algorithm is summarized in Algorithm \ref{algorithm2}. In addition, the unit-modulus constraint may not be satisfied by the obtained solution, thus the feasible IRS phase shifts are reconstructed by\begin{equation}\label{unit_modulus}
	v_{m}^{\text{opt}}=\frac{v_{m}}{\left|v_{m}\right|}, m \in \mathcal{M} .
\end{equation} The objective value of problem \eqref{prob2} is non-decreasing as the iterations proceed, and it is upper bounded by a finite value subject to
the limited budget power at the BS and the finite size of IRS \cite{convergence}. Therefore, the proposed algorithm terminates at the optimal objective value of \eqref{prob2}, which is a lower bound to that of problem \eqref{problem2}. The computational complexity of Algorithm \ref{algorithm2} is given by $\mathcal{O}\left(\left(9+M\right)^{3.5}\right)$ \cite{complexity}.

\section{Proposed Solution for the general system setup}

In this section, we study the general case with multiple DL and UL users. Compared to the previous simple case, the presence of MUI further complicates the optimization problem, and the SCA-based algorithm proposed above is not applicable. Thus, we propose an EW-based AO algorithm that decomposes \eqref{problem1} into two subproblems, and then alternatingly solve each subproblem until convergence. Specifically, for the IRS optimization subproblem, each phase shift is solved while others being fixed. The proposed algorithm is established on the observation that the WSR of the system can be recast as a difference of two concave functions.
Specifically, we can write $R=\alpha_{\text{U}}R_{\text{U}}+\alpha_{\text{D}}R_{\text{D}}=h\left(\mathbf{w},p,\mathbf{v}\right)-g\left(\mathbf{w},p,\mathbf{v}\right)$, where
\begin{align}
	h\left(\mathbf{w},p,\mathbf{v}\right)&\triangleq\alpha_{\text{D}}\sum_{k=1}^{K}\operatorname{log}\left(\sum_{ i=1}^{K}\left|\bar{\mathbf{h}}_{k}^H\mathbf{w}_i\right|^2+\sum_{l=1}^{L}p_{\text{U},l}\left|q_{l,k}\right|^2+\sigma_n^2\right)\nonumber \\
	&+\alpha_{\text{U}}\operatorname{log}\left|\sum_{l=1}^{L}p_{\text{U},l}\bar{\mathbf{f}}_{l}\bar{\mathbf{f}}_{l}^{H}+\sum_{k=1}^{K}\bar{\boldsymbol{G}}\mathbf{w}_k\mathbf{w}_k^H\bar{\boldsymbol{G}}^H+\sigma^2_n\mathbf{I}\right|\label{h},
\end{align}
\begin{align}
	g\left(\mathbf{w},p,\mathbf{v}\right)&\triangleq\alpha_{\text{D}}\sum_{k=1}^{K}\operatorname{log}\left(\sum_{i\neq k}^{K}\left|\bar{\mathbf{h}}_{k}^H\mathbf{w}_i\right|^2+\sum_{l=1}^{L}p_{\text{U},l}\left|q_{l,k}\right|^2+\sigma_n^2\right)\nonumber\\
	&+\alpha_{\text{U}}\operatorname{log}\left|\sum_{k=1}^{K}\bar{\boldsymbol{G}}\mathbf{w}_k\mathbf{w}_k^H\bar{\boldsymbol{G}}^H+\sigma^2_n\mathbf{I}\right|\label{g}.
\end{align}

\subsection{Optimizing Precoding Vector and User Transmit Power}
Denote $\mathbf{W}_k=\mathbf{w}_k\mathbf{w}_k^H, k\in\mathcal{K}$, and $\mathbf{W}=\{\mathbf{W}_k,k\in\mathcal{K}\}$. Then, with fixed $\mathbf{v}$, expressions \eqref{h} and \eqref{g} can be rewritten as
\begin{align}
	h\left(\mathbf{W},p\right)&=\alpha_{\text{D}}\sum_{k=1}^{K}\operatorname{log}_2\left(\sum_{i=1}^{K}\bar{\mathbf{h}}_{k}^H\mathbf{W}_i\bar{\mathbf{h}}_{k}+\sum_{l=1}^{L}p_{\text{U},l}\left|q_{l,k}\right|^2+\sigma_n^2\right)\nonumber\\
	&+\alpha_{\text{U}}\operatorname{log}_2\left|\sum_{l=1}^{L}p_{\text{U},l}\bar{\mathbf{f}}_{l}\bar{\mathbf{f}}_{l}^{H}+\sum_{k=1}^{K}\bar{\boldsymbol{G}}\mathbf{W}_k\bar{\boldsymbol{G}}^H+\sigma^2_n\mathbf{I}\right|,\\
	g\left(\mathbf{W},p\right)&=\alpha_{\text{D}}\sum_{k=1}^{K}\operatorname{log}_2\left(\sum_{i\neq k}^{K}\bar{\mathbf{h}}_{k}^H\mathbf{W}_i\bar{\mathbf{h}}_{k}+\sum_{l=1}^{L}p_{\text{U},l}\left|q_{l,k}\right|^2+\sigma_n^2\right)+\alpha_{\text{U}}\operatorname{log}_2\left|\sum_{k=1}^{K}\bar{\boldsymbol{G}}\mathbf{W}_k\bar{\boldsymbol{G}}^H+\sigma^2_n\mathbf{I}\right|.
\end{align}The subproblem is thus given as \begin{subequations}\label{subprob1}
	\begin{align}
		&\max_{\mathbf{W},p} ~h\left(\mathbf{W},p\right)-g\left(\mathbf{W},p\right)\\
		&~\text{s.t.} ~~ \sum_{k=1}^{K}\operatorname{Tr}\left(\mathbf{W}_k\right)\leq P_{\text{BS}}^{\text{max}},\label{W_constraint}\\
		&~\qquad \mathbf{W}_k\succeq \mathbf{0},k\in \mathcal{K},\label{33c}\\
		&~\qquad \text{rank}\left(\mathbf{W}_k\right)=1, k\in \mathcal{K}, \label{rank}\\
		&~\qquad \eqref{p_constraint}.
	\end{align}
\end{subequations}To find a solution, we start by applying the semidefinite relaxation (SDR) technique to get a relaxed version of \eqref{subprob1}, dropping the rank-1 constraints in \eqref{rank}. It is observed that both $h\left(\mathbf{W},p\right)$ and $g\left(\mathbf{W},p\right)$ are jointly concave w.r.t. $\mathbf{W}$ and $p$. Exploiting the fact that $\nabla_{\mathbf{X}} \log \left|\mathbf{I}+\mathbf{A} \mathbf{X} \mathbf{A}^H\right|=\mathbf{A}^H\left(\mathbf{I}+\mathbf{A} \mathbf{X} \mathbf{A}^H\right)^{-1} \mathbf{A}$ \cite{derivative}, we find the upper bound of $g\left(\mathbf{W},p\right)$, given the feasible points $\mathbf{W}^r$ and $p^r$, as follows,
	\begin{align}
		g\left(\mathbf{W},p\right)&\leq g\left(\mathbf{W}^r,p^r\right)+\alpha_{\text{D}}\sum_{k=1}^{K}\sum_{i\neq k}^{K}\left[\psi_k\left(\mathbf{W}^r,p^r\right)\bar{\mathbf{h}}_k^H\left(\mathbf{W}_i-\mathbf{W}_i^r\right)\bar{\mathbf{h}}_k\right]\nonumber\\
		&+\alpha_{\text{D}}\sum_{k=1}^{K}\sum_{l=1}^{L}\psi_k\left(\mathbf{W}^r,p^r\right)\left|q_{l,k}\right|^2\left(p_{\text{U},l}-p_{\text{U},l}^r\right)\nonumber\\
		&+\alpha_\text{U}\sum_{i=1}^{K}\operatorname{Tr}\left(\bar{\mathbf{G}}^H\boldsymbol{\Xi}\left(\mathbf{W}^r\right)\bar{\mathbf{G}}\left(\mathbf{W}_i-\mathbf{W}_i^r\right)\right)\triangleq g^{\text{ub}}\left(\mathbf{W},p\right),
	\end{align}
where $\psi_k\left(\mathbf{W}^r,p^r\right)\!=\!\left(\sum_{m\neq k}^{K}\bar{\mathbf{h}}_{k}^H\mathbf{W}_m^r\bar{\mathbf{h}}_{k}\!+\!\sum_{l=1}^{L}p_{\text{U},l}^r\left|q_{l,k}\right|^2\!+\!\sigma_n^2\right)^{-1}$ and  $\text{ }\!\boldsymbol{\Xi}\left(\mathbf{W}^r\right)\!=\!\left(\sum_{k=1}^{K}\bar{\boldsymbol{G}}\mathbf{W}^r_k\bar{\boldsymbol{G}}^H\!\right. \\
\left.+\sigma^2_n\mathbf{I}\right)^{-1}$.

Then, subproblem in \eqref{subprob1} is approximated as
\begin{subequations}\label{power_subprob}
	\begin{align}
		&\max_{\mathbf{W},p}~ h\left(\mathbf{W},p\right)-g^{\text{ub}}\left(\mathbf{W},p\right)\\
		&~\text{s.t.} ~~ \eqref{p_constraint},\eqref{W_constraint},\eqref{33c},
	\end{align}
\end{subequations}which is convex and is readily to be solved by existing solvers such as CVX \cite{cvx}. Based on the rank reduction result in Theorem 3.2 of \cite{rank}, the optimal solution $\mathbf{W}$ to the SDR problem \eqref{power_subprob} satisfies the rank-1 constraints. Thus, the adopted SDR is tight for problem \eqref{power_subprob}.


\subsection{Optimizing Passive Beamforming}
For any given precoding matrices $\mathbf{W}$ and UL user power allocation $p$, the IRS phase shift optimization in problem \eqref{problem1} can be simplified as\begin{subequations}\label{subproblem2}
	\begin{align}
		&\max_{\mathbf{v}} ~h\left(\mathbf{v}\right)-g\left(\mathbf{v}\right)\label{obj2}\\
		&~\text{s.t.} ~~ \eqref{constraint3}.
	\end{align}
\end{subequations}
The challenges for solving subproblem \eqref{subproblem2} lie in the non-concave objective function \eqref{obj2} and the non-convex unit-modulus constraint on each reflection coefficient in \eqref{constraint3}. To tackle them, an EW-based algorithm is proposed, where the rationale is to optimize the IRS phase shifts in a one-by-one manner. Since the dependence of objective \eqref{obj2} on the phase shifts is implicit, the obstacle here is to derive an explicit function of each phase shift such that it can be efficiently optimized. Let  $\mathbf{G}_t=\left[\mathbf{g}_{t,1},\cdots,\mathbf{g}_{t,M}\right]^H$ and $\mathbf{h}_{r,k}^H=\left[h_{r,k,1}^H,\cdots,h_{r,k,M}^H\right]$,  where $\mathbf{g}_{t,m}\in\mathbb{C}^{N_{\text{T}}\times 1}$ is the $m$-th column of $\mathbf{G}_t$ and $h_{r,k,m}^H$ is the $m$-th element of $\mathbf{h}_{r,k}^H$. Thus, the effective DL channel $\bar{\mathbf{h}}_k^H$ can be rewritten as\begin{equation}
	\bar{\mathbf{h}}_k^H=\mathbf{h}_{d,k}^H+\sum_{m=1}^{M}v_{m}h_{r,k,m}^H\mathbf{g}_{t,m}^H.
\end{equation}
Then, we can rewrite $\left|\bar{\mathbf{h}}_k^H\mathbf{w}_i\right|^2$ as 
\begin{align}\label{33}
	\left|\bar{\mathbf{h}}_k^H\mathbf{w}_i\right|^2
	&=\tilde{\mathbf{h}}_{k,m}^H\mathbf{W}_i\tilde{\mathbf{h}}_{k,m}+v_{m}h_{r,k,m}^*\mathbf{g}_{t,m}^H\mathbf{W}_{i}\tilde{\mathbf{h}}_{k,m}+v_{m}^*h_{r,k,m}\tilde{\mathbf{h}}_{k,m}^H\mathbf{W}_i\mathbf{g}_{t,m}\nonumber\\
	&+\left|h_{r,k,m}\right|^2\mathbf{g}_{t,m}^H\mathbf{W}_i\mathbf{g}_{t,m}
	\triangleq f_{k,i}\left(v_m\right),
\end{align}where $\tilde{\mathbf{h}}_{k,m}^H=\mathbf{h}_{d,k}^H+\sum_{j\neq m}^{M}v_{j}h_{r,k,j}^H\mathbf{g}_{t,j}^H$. The equality in \eqref{33} holds since $|v_m|=1$. Similarly, denote $\mathbf{G}_r=\left[\mathbf{g}_{r,1},\cdots,\mathbf{g}_{r,M}\right]$ and  $\mathbf{f}_{r,l}=\left[f_{r,l,1},\cdots,f_{r,l,M}\right]^T$, where $\mathbf{g}_{r,m}\in\mathbb{C}^{N_R\times 1}$ is the  $m$-th column of $\mathbf{G}_r$ and $f_{r,l,m}$ is the $m$-th element of $\mathbf{f}_{r,l}$. As such, the effective SI channel and UL channel can thus be respectively re-expressed as\begin{equation}
	\bar{\boldsymbol{G}}=\mathbf{G}_{SI}+\sum_{m=1}^{M}v_{m}\mathbf{g}_{r,m}\mathbf{g}_{t,m}^H,
\end{equation}\begin{equation}
	\bar{\mathbf{f}}_{l}=\mathbf{f}_{d,l}+\sum_{m=1}^{M}v_{m}\mathbf{g}_{r,1}f_{r,l,m}.
\end{equation}
Thus, we can rewrite $\bar{\boldsymbol{G}}\mathbf{w}_k\mathbf{w}_k^H\bar{\boldsymbol{G}}^H$ and $\bar{\mathbf{f}}_{l}\bar{\mathbf{f}}_l^H$ as
\begin{align}
	\bar{\boldsymbol{G}}\mathbf{w}_k\mathbf{w}_k^H\bar{\boldsymbol{G}}^H&=\tilde{\mathbf{G}}_{m}\mathbf{W}_k\tilde{\mathbf{G}}_{m}^H+v_m\mathbf{g}_{r,m}\mathbf{g}_{t,m}^H\mathbf{W}_{k}\tilde{\mathbf{G}}_{m}^H+v_m^*\tilde{\mathbf{G}}_{m}\mathbf{W}_k\mathbf{g}_{t,m}\mathbf{g}_{r,m}^H\nonumber\\ &+\mathbf{g}_{r,m}\mathbf{g}_{t,m}^H\mathbf{W}_{k}\mathbf{g}_{t,m}\mathbf{g}_{r,m}^H
	\triangleq \hat{\mathbf{F}}_{k}\left(v_m\right),\\
	\bar{\mathbf{f}}_{l}\bar{\mathbf{f}}_l^H&=\tilde{\mathbf{f}}_{l,m}\tilde{\mathbf{f}}_{l,m}^H+v_{m}f_{r,l,m}\mathbf{g}_{r,m}\tilde{\mathbf{f}}_{l,m}^H+v_{m}^*f_{r,l,m}^*\tilde{\mathbf{f}}_{l,m}\mathbf{g}_{r,m}^H+\left|f_{r,l,m}\right|^2\mathbf{g}_{r,m}\mathbf{g}_{r,m}^H
	\triangleq \bar{\mathbf{F}}_{l}\left(v_m\right),
\end{align}where $\tilde{\mathbf{G}}_{m}=\left(\mathbf{G}_{SI}+\sum_{j\neq m}^{M}v_{j}\mathbf{g}_{r,j}\mathbf{g}_{t,j}^H\right)$ and $\tilde{\mathbf{f}}_{l,m}=\left(\mathbf{f}_{d,l}+\sum_{j\neq m}^{M}v_{j}\mathbf{g}_{r,j}f_{r,l,j}\right)$.
Then, $h\left(v_m\right)$ and $g\left(v_m\right)$ become
\begin{align}
	h\left(v_m\right)&=\alpha_{\text{D}}\sum_{k=1}^{K}\operatorname{log}_2\left(\sum_{i=1}^{K}f_{k,i}\left(v_m\right)+\sum_{l=1}^{L}p_{U,l}\left|q_{l,k}\right|^2+\sigma_n^2\right)\nonumber\\
	&+\alpha_{\text{U}}\operatorname{log}_2\left|\sum_{l=1}^{L}p_{U,l}\bar{\mathbf{F}}_{l}\left(v_m\right)+\sum_{k=1}^{K}\hat{\mathbf{F}}_{k}\left(v_m\right)+\sigma^2_n\mathbf{I}\right|, \label{20}\\
	g\left(v_m\right)&=\alpha_{\text{D}}\sum_{k=1}^{K}\operatorname{log}_2\left(\sum_{i\neq k}^{K}f_{k,i}\left(v_m\right)+\sum_{l=1}^{L}p_{U,l}\left|q_{l,k}\right|^2+\sigma_n^2\right)+\alpha_{\text{U}}\operatorname{log}_2\left|\sum_{k=1}^{K}\hat{\mathbf{F}}_{k}\left(v_m\right)+\sigma^2_n\mathbf{I}\right|\nonumber\\
	&=\alpha_{\text{D}}\sum_{k=1}^{K}\operatorname{log}_2\left(\sum_{i\neq k}^{K}f_{k,i}\left(v_m\right)+\sum_{l=1}^{L}p_{U,l}\left|q_{l,k}\right|^2+\sigma_n^2\right)+g^\prime\left(v_m\right).\label{g_v}
\end{align}It can be readily checked that functions $h\left(v_m\right)$ and $g\left(v_m\right)$ are both concave w.r.t. $v_m$, and it is straightforward to apply SCA on $g\left(v_m\right)$. However, in order to reduce complexity, we perform some manipulations on $g\left(v_m\right)$ before applying SCA \cite{zsw}. Let $ \mathbf{A}_m=\sigma_n^2\mathbf{I}+\tilde{\mathbf{G}}_{m}\mathbf{W}\tilde{\mathbf{G}}_{m}^H+\mathbf{g}_{r,m}\mathbf{g}_{t,m}^H\mathbf{W}\mathbf{g}_{t,m}\mathbf{g}_{r,m}^H$ and $ \mathbf{B}_m=\mathbf{g}_{r,m}\mathbf{g}_{t,m}^H\mathbf{W}\tilde{\mathbf{G}}_{m}^H$. Then,
\begin{subequations}
	\begin{align}
		g^\prime\left(v_m\right)=& \log \left|\sum_{k=1}^{K} \mathbf{\hat{F}}_k\left(v_m\right)+\sigma_n^2\mathbf{I}\right| \\
		=&\log_2 \operatorname{det}\left(\mathbf{A}_m+v_m \mathbf{B}_m+v_m^* \mathbf{B}_m^H\right) \\
		=&\log_2 \operatorname{det} \mathbf{A}_m\left(\mathbf{I}+v_m \mathbf{A}_m^{-1} \mathbf{B}_m+v_m^* \mathbf{A}_m^{-1} \mathbf{B}_m^H\right)\\
		=&\log_2 \operatorname{det}\mathbf{A}_m+\log \operatorname{det}  \left(\mathbf{I}+v_m \mathbf{A}_m^{-1} \mathbf{B}_m+v_m^* \mathbf{A}_m^{-1} \mathbf{B}_m^H\right)\\
		\triangleq &\log_2 \operatorname{det}\mathbf{A}_m+g^{\prime\prime}\left(v_m\right).
	\end{align}
\end{subequations}
Due to the fact that $\operatorname{rank}(\mathbf{B}_m) = 1$, the matrix $\mathbf{A}_m^{-1}\mathbf{B}_m$ is also rank one. Next, we study the following two cases, in which $\mathbf{A}_m^{-1}\mathbf{B}_m$ is either diagonalizable or not, respectively.

\emph{Case I: Diagonalizable $\mathbf{A}_m^{-1} \mathbf{B}_m$:} First, we study the case where $\mathbf{A}_m^{-1} \mathbf{B}_m$ is diagonalizable and hence its eigenvalue decomposition (EVD) exists. Since $\operatorname{rank}\left(\mathbf{A}_m^{-1}\mathbf{B}_m\right)=1$, we can express its EVD as
$\mathbf{A}_m^{-1}\mathbf{B}_m=\mathbf{U}_m \mathbf{\Sigma}_m \mathbf{U}_m^{-1}$, where $ \mathbf{U}_m \in \mathbb{C}^{N_{\text{R}} \times N_{\text{R}}}$ and $\mathbf{\Sigma}_m=\operatorname{diag}\left\{\lambda_m, 0, \dots, 0 \right\} \in \mathbb{C}^{N_{\text{R}} \times N_{\text{R}}}$, with $\lambda_m\in\mathbb{C}$ denoting the only non-zero eigenvalue. Then, $g^{\prime\prime}\left(v_m\right)$ is given by
\begin{subequations}
	\begin{align}
		g^{\prime\prime}\left(v_m\right)
		&=\log_2  \operatorname{det}\left(\mathbf{I}+v_m \mathbf{\Sigma}_m+v_m^* \mathbf{U}_m^{-1} \mathbf{A}_{m}^{-1}\left(\mathbf{U}_m^{-1}\right)^H \mathbf{\Sigma}_m^H \mathbf{U}_m^H \mathbf{A}_m^H \mathbf{U}_m\right)\\
		&=\log_2 \operatorname{det}\left(\mathbf{I}+v_m \mathbf{\Sigma}_m+v_m^* \mathbf{R}_m^{-1} \mathbf{\Sigma}_m^H \mathbf{R}_m\right).
	\end{align}
\end{subequations}
Let $\mathbf{r}_m\in\mathbb{C}^{N_{\text{R}\times 1}}$ denote the first column of $\mathbf{R}_m^{-1}$ and $\mathbf{r}'^{T}_m\in\mathbb{C}^{1\times N_{\text{R}}}$ the first row of $\mathbf{R}_m$. Then, we have
\begin{subequations}
	\vspace{-0.2cm}
	\begin{align}
		g^{\prime \prime}\left(v_m\right)& =\log_2 \operatorname{det}\left(\mathbf{I}+v_m \mathbf{\Sigma}_m+v_m^* \mathbf{r}_m \lambda_m^* \mathbf{r}_m^{\prime T}\right) \\
		&=\log_2\left(1+\left|\lambda_m\right|^2\left(1-r_{m,1}^{\prime} r_{m,1}\right)+2 \left(v_m \lambda_m+v_m^* \lambda_m^*\right)\right).
	\end{align}
\end{subequations}Though $g\left(v_m\right)=\alpha_{\text{U}}\left(g^{\prime \prime}\left(v_m\right)+\log\operatorname{det}\mathbf{A}_m\right)+\alpha_{\text{D}}\sum_{k=1}^{K}\operatorname{log}\left(\sum_{i\neq k}^{K}f_{k,i}\left(v_m\right)+\sum_{l=1}^{L}p_{\text{U},l}\left|q_{l,k}\right|^2+\sigma_n^2\right)$ has been simplified, the objective function $h(v_m)-g(v_m)$ is still non-concave. Therefore, we derive its upper bound by taking the first-order Taylor expansion around any feasible point $v^r_m$, as\begin{align}
		g\left(v_m\right)&\leq g\left(v_m^r\right)+
		\alpha_D\sum_{k=1}^{K}\operatorname{Tr}\left(\gamma_k\left(v_m^r\right)\sum_{i\neq k}^{K}h_{r,k,m}^*\mathbf{g}_{t,m}^H\mathbf{W}_{i}\tilde{\mathbf{h}}_{k,m}\left(v_m-v_m^r\right)\right)\nonumber\\
		&+\alpha_D\sum_{k=1}^{K}\operatorname{Tr}\left(\gamma_k\left(v_m^r\right)\sum_{i\neq k}^{K}h_{r,k,m}\tilde{\mathbf{h}}_{k,m}^H\mathbf{W}_i\mathbf{g}_{t,m}\left(v_m^*-v_m^{r,*}\right)\right)\nonumber\\
		&+2\alpha_U\lambda_m\eta\left(v_m^r\right)\left(v_m-v_m^r\right)+2\alpha_U\lambda^*_m\eta\left(v_m^r\right)\left(v_m^*-v_m^{r,*}\right)\triangleq g^{\text{ub}}\left(v_m\right),
	\end{align}
where $\gamma_k\left(v_m^r\right)=\left(\sum_{i\neq k}^{K}f_{k,i}\left(v_m^r\right)+\sum_{l=1}^{L}p_{U,l}\left|q_{l,k}\right|^2+\sigma_n^2\right)^{-1}$ and 
$\eta\left(v_m^r\right)=\left(1+\left|\lambda_m\right|^2\left(1-r_{m,1}^{\prime} r_{m,1}\right)\right. \\
\left.+2 \left(v_m^r \lambda_m+v_m^{r,*} \lambda_m^*\right)\right)^{-1}$. Then, this subproblem is given by \begin{subequations}\label{v_subprob}
	\begin{align}
		&\max_{v_m}~ h\left(v_m\right)-g^{\text{ub}}\left(v_m\right)\\
		&~\text{s.t.}~~ |v_m|=1\label{unit_modulus_v_m}.
		\vspace{-0.5cm}
	\end{align}
\end{subequations}Finally, for the non-convex constraint \eqref{unit_modulus_v_m}, since the unit-modulus property is assumed in the above manipulations, the relaxation technique of the previous section cannot be applied. Thus, we apply the square penalty approach \cite{penalty} by adding a term $\rho|v_m|^2$ into the objective function in \eqref{v_subprob}, where  $\rho\gg0$ is a defined sufficiently large penalty factor exploited to ensure the unit-modulus constraint at the optimal solution. The advantage is that such equivalence only needs a large $\rho$, thus it does not need to be adjusted gradually. For the non-concave term $\rho|v_m|^2$, a lower bound for $|v_m|^2$ can be derived by SCA. Specifically, for any given $v_m^r$, we have \begin{equation}\label{45}
|v_m|^2 \geq-\left|v_m^r\right|^2+2 \operatorname{Re}\left\{v_m^* v_m^r\right\},
\end{equation}which is an affine function w.r.t. $v_m$.

Consequently, based on \eqref{45} and dropping the irrelevant terms, the subproblem is approximated as \looseness=-1 \begin{subequations}\label{subproblem2_2}
\begin{align}
	&\max_{v_m}~ h\left(v_m\right)-g^{\text{ub}}\left(v_m\right)+2\rho\operatorname{Re}\left\{v_m^* v_m^r\right\}\\
	&~\text{s.t.} ~~ |v_m|\leq 1\label{relaxed_EW}.
\end{align}
\end{subequations}Problem \eqref{subproblem2_2} is shown to be convex and can be solved by off-the-shelf tools such as CVX \cite{cvx}.

\emph{Case II: Non-Diagonalizable $\mathbf{A}_m^{-1} \mathbf{B}_m$:} If $\mathbf{A}_m^{-1}\mathbf{B}_m$ is non-diagonalizable, $g''\left(v_m\right)$ is independent of $v_m$ \cite{zsw}, which thus can be ignored during the optimization of $v_m$. Thus, this subproblem is given by\vspace{-0.3cm} \begin{subequations}\label{v_subprob2}
	\begin{align}
		&\max_{v_m}~ h\left(v_m\right)-\tilde{g}^{\text{ub}}\left(v_m\right)+2\rho\operatorname{Re}\left\{v_m^* v_m^r\right\}\\
		&~\text{s.t.}~~ \eqref{relaxed_EW},
	\end{align}
\vspace{-0.5cm}
\end{subequations}
where 	\begin{align}
		\tilde{g}^{\text{ub}}\left(v_m\right)&=g\left(v_m^r\right)+\alpha_{\text{D}}\sum_{k=1}^{K}\operatorname{Tr}\left(\gamma_k\left(v_m^r\right)\sum_{i\neq k}^{K}h_{r,k,m}^*\mathbf{g}_{t,m}^H\mathbf{W}_{i}\tilde{\mathbf{h}}_{k,m}\left(v_m-v_m^r\right)\right)\nonumber\\
		&+\alpha_{\text{D}}\sum_{k=1}^{K}\operatorname{Tr}\left(\gamma_k\left(v_m^r\right)\sum_{i\neq k}^{K}h_{r,k,m}\tilde{\mathbf{h}}_{k,m}^H\mathbf{W}_i\mathbf{g}_{t,m}\left(v_m^*-v_m^{r,*}\right)\right).
	\end{align}
Again, problem \eqref{v_subprob2} is convex and can be readily solved by existing solvers.

\subsection{Overall Algorithm and Computational Complexity}
\begin{algorithm}[t]
	\caption{EW-based Algorithm for Solving Problem \eqref{problem1}}\label{algorithm}
	\begin{algorithmic}[1]
		\State \textbf{Initialize} the IRS phase-shift vector $\mathbf{v}^r$, DL BS precoding matrices $\{\mathbf{W}_k^r\}$, UL user power $\{p_{\text{U},l}^r\}$, $\rho$, iteration index $r=0$, and threshold $\epsilon$.
		\Repeat \label{step2}
		\State Update the DL BS transmit beamforming and the UL user power by solving problem \eqref{power_subprob}.
		\Loop \textbf{ for} $m$
		\If{$\mathbf{A}_m^{-1}\mathbf{B}_m$ is diagonalizable} 
		\State Update the IRS phase shift $v_m$ by solving problem \eqref{subproblem2_2}.
		\Else
		\State Update the IRS phase shift $v_m$ by solving problem \eqref{v_subprob2}.
		\EndIf
		\EndLoop
		\Until the increase of the objective in problem \eqref{problem1} is below $\epsilon$. \label{step11}
	\end{algorithmic}
\end{algorithm}

The overall algorithm is sumarized in Algorithm \ref{algorithm} based on the above solutions to the  subproblems. It is guaranteed that the proposed AO algorithm will be converged to a point since the limited transmit power at the BS and the UL users constrains the objective of \eqref{problem1} to be finite. In addition, the objective value of problem \eqref{problem1} is non-decreasing with each iteration due to the fact that each subproblem is solved locally and/or optimally.

 The main computational complexity lies in steps 2 to 9 of Algorithm \ref{algorithm}. Specifically, in step 3, problem \eqref{power_subprob} has $K+L$ variables and constraint in \eqref{W_constraint} is of dimension $N_{\text{T}}^2$. Thus, the complexity for solving problem \eqref{power_subprob} is $\mathcal{O}\left(\left(K+L\right)\left(N_{\text{T}}^2\right)\right)^{3.5}$ \cite{complexity}. In step 6, the complexity mainly comes from the matrix calculations. Since \eqref{subproblem2_2} involves a logarithmic form, its complexity is  given by $\mathcal{O}\left(2^{3.5}\left(N_{\text{R}}^3\right)\right)$. Consequently, the overall complexity of Algorithm \ref{algorithm} is $\mathcal{O}\left(N_{\text{iter}}\left(\left(\left(K+L\right)\left(N_{\text{T}}^2\right)\right)^{3.5}\right. \right.\\
 \left.\left.+2^{3.5}M\left(N_{\text{R}}^3\right)\right)\right)$, where $N_{\text{iter}}$ represents the total number of iterations required for convergence.

	\section{Numerical Results}
	In this section, numerical results are presented to demonstrate the effectiveness of the proposed algorithm, which also offer insights for the IRS-assisted FD system. We consider a uniform linear array at the BS and a uniform rectangular array at the IRS along the $x$-axis and in the $x$-$y$ plane of a three-dimensional (3D) coordinate system, measured in meter (m). The centers of the BS and IRS arrays are respectively located at $(0,0,3)$ m and $(0,0,5.5)$ m. Any two IRS elements are separated by $\lambda/2$. The distributions of UL and DL users are uniform and random in two respective circles, with the centers at $(15,0,1)$ and $(25,0,1)$ m, and a radius of $5$ m. Furthermore, it is assumed that the separation between transmit antennas and receive antennas at the BS is $0.2$ m \cite{loop_channel}. The LI channel of the BS, i.e., $\mathbf{G}_{\text{SI}}$, is modeled as $\mathcal{C} \mathcal{N}_{N_{\mathrm{R}} N_{\mathrm{T}}}\left(\sqrt{\sigma_{\mathrm{SI}}^2 K /(1+K)} \bar{\mathbf{H}}_{\mathrm{SI}},\left(\sigma_{\mathrm{SI}}^2 /(1+K)\right) I_{N_{\mathrm{R}}} \otimes I_{N_{\mathrm{T}}}\right)$, where $K=30$ dB and $\bar{\mathbf{H}}_{\mathrm{SI}}$ is a matrix of all ones \cite{spectral_efficiency}. 
	
	To suppress the SI at the BS, it is preferable to deploy the IRS close to the BS. We assume that the BS is within the Frauhofer distance  $R_f=\frac{2L_{\text{IRS}}^2}{\lambda}$ m of the IRS, where $L_{\text{IRS}}$ is the maximum physical dimension of the IRS. Thus, in our model, the BS is located  in the radiating near field of the IRS \cite{pathloss}, and considering the spherical nature of the transmitted waves, the $(M_x,M_y)$-th entry of the LoS channel between the IRS and the $i$-th transmit antenna of the BS can be modeled as \cite{near_field} \begin{equation}
		[\mathbf{M}_i^t]_{M_x,M_y}=\text{PL}^t_{i,M_x,M_y}e^{-j\frac{2\pi}{\lambda}d^t_{i,M_x,M_y}},
	\end{equation}where $\text{PL}^t_{i,M_x,M_y}$ is the corresponding free space pathloss model: \begin{equation}\label{48}
		\text{PL}^t_{i,M_x,M_y}=\frac{c}{4 \pi f d_{i,M_x,M_y}^{t}}.
\end{equation}In \eqref{48}, $c$ is the speed-of-light, $f$ is the carrier frequency, and $d_{i,M_x,M_y}^{t}$ is the link distance. Therefore, the baseband equivalent channel between the BS transmist antennas and the IRS is given by  $\mathbf{G}_t=[\operatorname{vec}(\mathbf{M}^t_1),\dots,\operatorname{vec}(\mathbf{M}^{t}_{N_{\text{T}}})]$. The baseband equivalent channel between the BS receive antennas and the IRS, i.e., $\mathbf{G}_r$, can be obtained in a similar manner. The other channels are assumed to be in the far field. We apply a distance-dependent path loss model in which $L(d)=c_0\left(d / d_0\right)^{-\alpha}$, where $c_0=(\lambda /(4 \pi))^2$ is the path loss at the reference distance $d_0=1$ m, $d$ denotes the link distance, and $\alpha$ denotes the path loss exponent. Considering the small-scale fading, we model the IRS-user link, the BS-user link, and the links between users to follow the Rician fading with the Rician factor being $3$ dB, and the path-loss exponents are set to be $2.2$, $3.6$, and $3.6$, respectively. The angles of arrival and departure  are assumed to be randomly distributed over the range $[0, 2\pi)$. We assume that power budget is the same for all UL users, i.e., $p_{\text{U}}^{\text{max}}$. Unless otherwise specified, we set $f=2.4$ GHz, $P_{\text{BS}}^{\text{max}}=26$ dBm, $p_{\text{U}}^{\text{max}}=23$ dBm, $\sigma_{n}^2=-80$ dBm, $\alpha_{\text{D}}=\alpha_{\text{U}}=0.5$, $\rho=100$, and $\epsilon=10^{-3}$.

	\subsection{System Setup with One DL and One UL User}
	To demonstrate the performance gain offered by the proposed scheme, we compare it with the following methods in the simple case: 1) \textbf{Proposed}: The solutions are obtained by Algorithm \ref{algorithm2}; 2) \textbf{SIC}: The IRS phase shifts are designed to minimize the f-norm of the SI channel; With the obtained IRS phase shifts, the BS power and the UL power are optimized by Algorithm \ref{algorithm2}\footnote{The SIC-based scheme is proceeded without performing the reconstruction such that a performance upper bound can be achieved for better comparison.}; 3) \textbf{DL-assisted}: The IRS phase shifts are designed to maximize the DL rate by assuming that there are no UL users; With the obtained IRS phase shifts, the BS power and the UL power are then optimized by Algorithm \ref{algorithm2}; 4) \textbf{UL-assisted}: The IRS phase shifts are designed to maximize the UL rate by assuming that there are no DL users; With the obtained IRS phase shifts, the BS power and the UL power are then optimized by Algorithm \ref{algorithm2}; 5) \textbf{Random IRS}: Only the BS power and the UL power are optimized by Algorithm \ref{algorithm2}, with each phase shift randomly chosen over the range $[0,2 \pi)$; 6) \textbf{Without IRS}: Shows performance achieved without an IRS.
		
	To begin with, we analyze the simple case to investigate how the IRS cancels SI without the presence of MUI and CCI.
	
	\begin{figure}
		\vspace{-0.5cm}
		\centering
		\includegraphics[width=.5\linewidth]{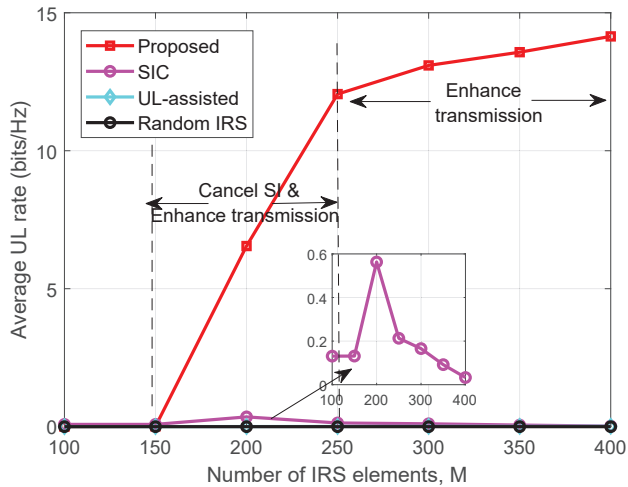}
		\vspace{-0.5cm}
		\caption{Average UL rate versus the number of IRS elements with $\sigma^2_{\text{SI}}=-40\text{ dBW}$.}
		\label{single_UL_rate_vs_N}
		\vspace{-0.5cm}
	\end{figure}

\subsubsection{SI cancellation capability of the IRS}Before studying the WSR of the system, it is noticed that the SI only restricts the UL rate. Hence, to investigate the SI cancellation capability of the IRS, we first focus on the UL transmission.

In Fig. \ref{single_UL_rate_vs_N}, we study the average UL rate versus $M$ with $\sigma_{\text{SI}}^2=-40 \text{ dBW}$, $P_{\text{BS}}= P_{\text{BS}}^{\text{max}}$, and $p_{\text{U}}=p_{\text{U}}^{\text{max}}$. First, it is observed that, as expected, the proposed scheme outperforms all benchmarks since the IRS is capable of balancing the SI mitigation and transmission enhancement to improve the system performance. In addition, it is observed that the performance of the proposed approach remains static at first when $M$ is small, followed by a dramatic increase as $M$ increases from $150$ to $250$, and with diminishing returns thereafter. This behavior can be attributed to the fact that when $M$ is small, the magnitude of the reflected channel is not comparable to that of the LI channel, and thus the SI cannot be significantly suppressed, resulting in negligible performance improvement. On the other hand, with $150\leq M\leq250$, the IRS is able to simultaneously suppress the SI and enhance the UL transmission, accounting for the dramatic increase in the UL rate. For $M\geq 250$, since the SI is fully mitigated, the gain in UL rate mainly comes from the passive beamforming gain in improving information transmission quality. Thus, the gain exhibits a comparatively steady pace. Similar behavior can also be observed for the SIC-based approach, where the UL rate saturates for $M \geq 250$. This also indicates that the SIC-based scheme cannot enhance transmission. Finally, it can be observed that the UL-assisted scheme performs similarly to the scheme with random IRS. It is easy to understand why the latter has sn unsatisfactory performance. However, for the former, the poor performance is because the UL rate suffers severely from the strong SI, even if the IRS is designed to improve transmission efficiency.

\begin{figure}
	\centering
	\includegraphics[width=.5\linewidth]{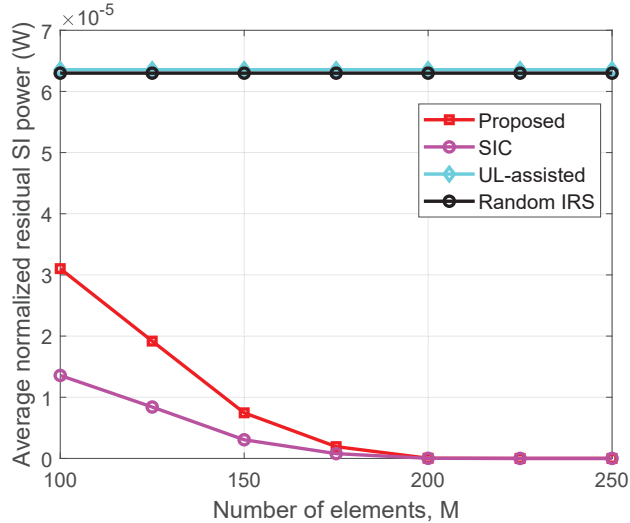}
	\vspace{-0.4cm}
	\caption{Average normalized residual SI power versus the number of IRS elements with $\sigma^2_{\text{SI}}=-40\text{ dBW}$.}
	\label{single_UL_rate_vs_SI2}
			\vspace{-0.5cm}
\end{figure}

To visualize the capability of IRS in SI mitigation, we plot in Fig. \ref{single_UL_rate_vs_SI2} the average normalized residual SI power $\bar{\mathbf{G}}$ versus $M$. It can be seen that the residual SI power with the proposed scheme and the SIC-based scheme decreases to zero. This thus verifies the above explanation that when $M$ is large, the SI can be perfectly mitigated with these two schemes. Moreover, the UL-assisted scheme and that without IRS cannot suppress the strong SI. Thus, this again explains why the UL rates achieved by these two approaches are almost zero.

\begin{figure}
	\vspace{-0.5cm}
	\centering
	\includegraphics[width=.5\linewidth]{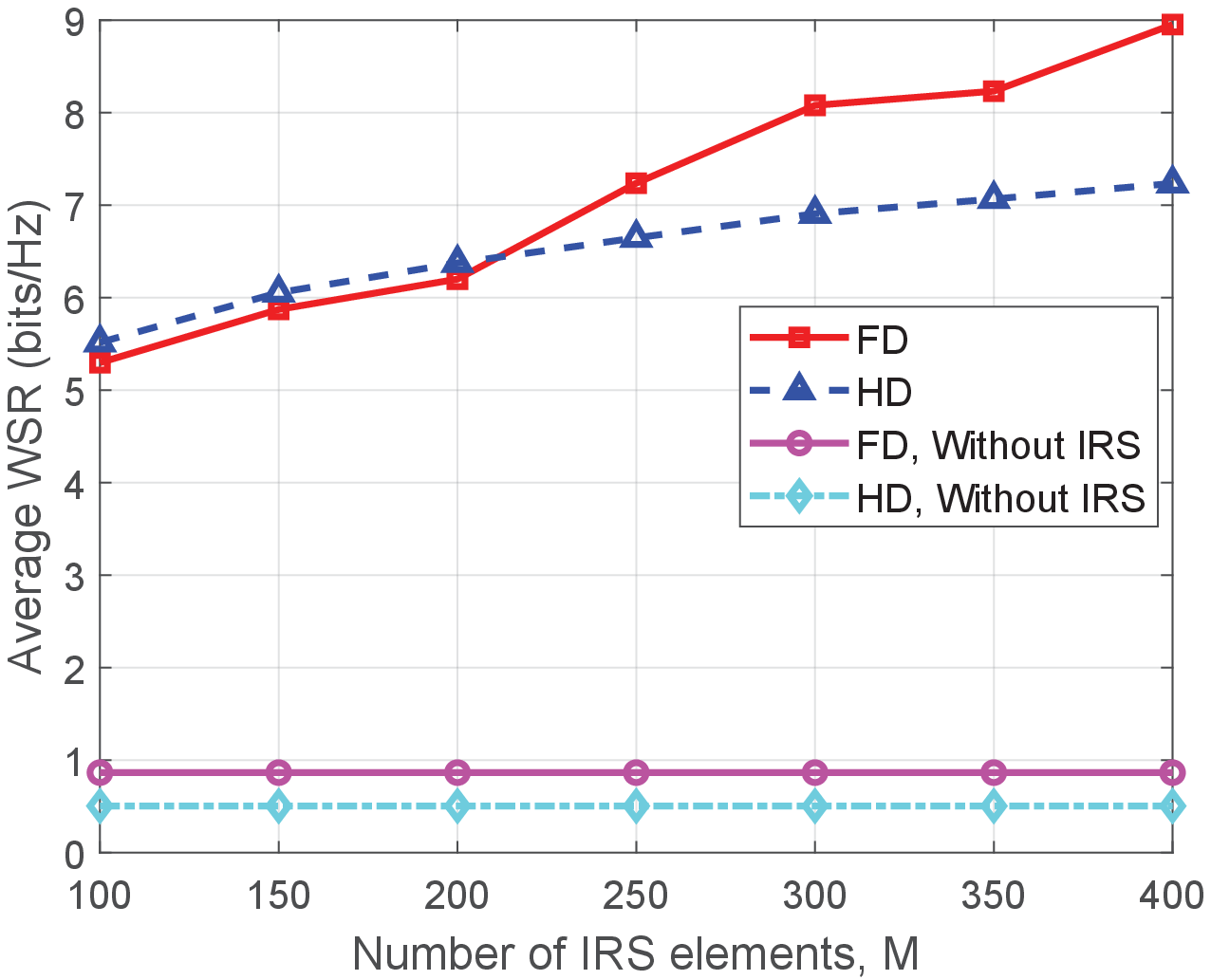}
	\vspace{-0.5cm}
	\caption{Average WSR versus the number of IRS elements  with $\sigma^2_{\text{SI}}=-40\text{ dBW}$.}
	\label{single_UL_DL_rate_vs_N1}
\end{figure}
\begin{figure}
\centering
\includegraphics[width=.5\linewidth]{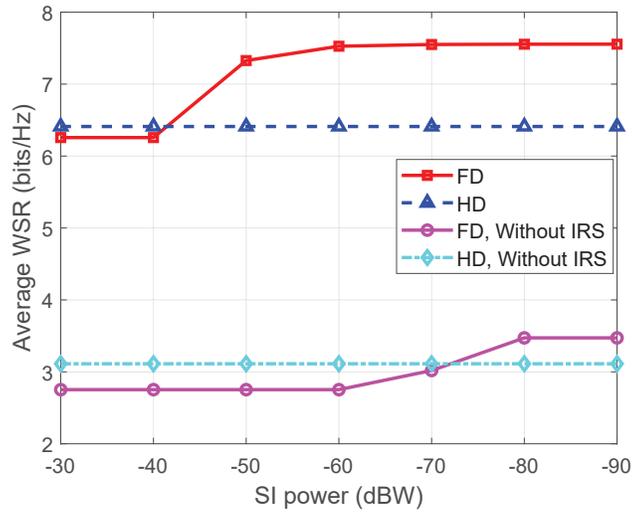}
\vspace{-0.5cm}
\caption{Average WSR versus the SI power with $M=200$.}
\label{single_UL_DL_rate_vs_SI1}
		\vspace{-0.5cm}
\end{figure}

	\subsubsection{FD system versus HD system} 
\begin{figure}
	\vspace{-0.5cm}
	\centering
	\includegraphics[width=.5\linewidth]{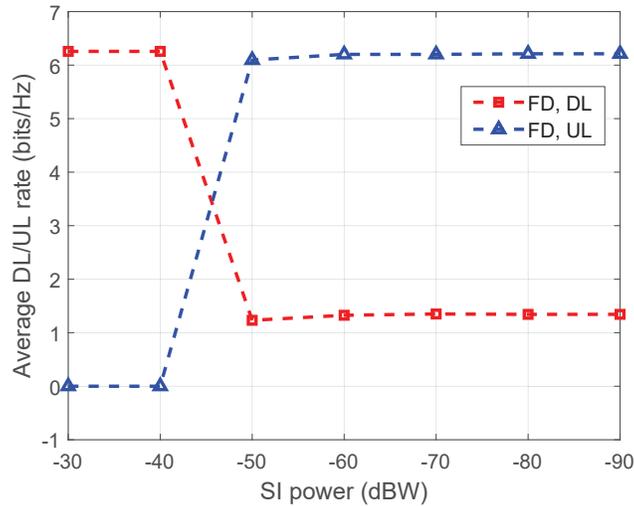}
	\vspace{-0.5cm}
	\caption{Average DL/UL rate versus the SI power for the proposed algorithm.}
	\label{single_UL_DL_rate_vs_SI3}
\end{figure}

\begin{figure}	
	\centering
	\includegraphics[width=.5\linewidth]{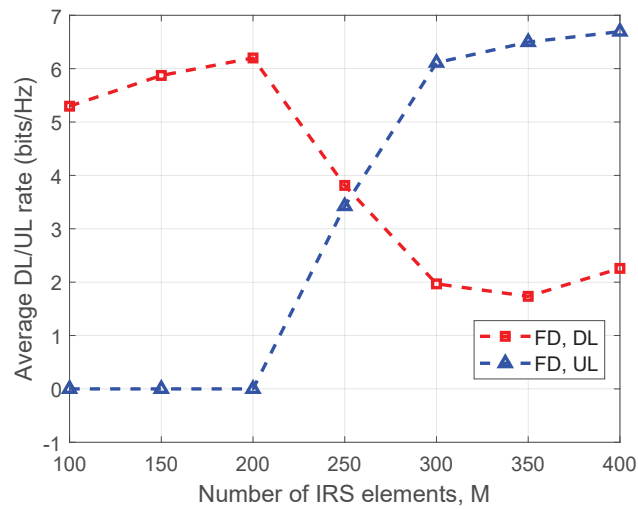}
	\vspace{-0.5cm}
	\caption{Average DL/UL rate versus the number of IRS elements for the proposed algorithm.}
	\label{single_UL_DL_rate_vs_N2}
		\vspace{-0.5cm}
\end{figure}

In Fig. \ref{single_UL_DL_rate_vs_N1}, we study the FD and HD system performance in terms of average WSR to learn the importance of SI cancellation. We consider the following approaches for comparison: 1) \textbf{FD}: Our proposed algorithm; 2) \textbf{HD}: The BS operates in HD rather than FD mode; 3) \textbf{FD, without IRS}: It is the proposed algorithm but without an IRS; 4) \textbf{HD, without IRS}: Same as (3) except the BS operates in HD mode. Firstly, it is shown that the WSR obtained by all approaches  monotonically increases with $M$ since additional IRS elements provide a larger passive beamforming gain for both the DL and UL transmissions. In addition, it can be seen that with a small and hence strong SI, the FD system is worse than the HD system in terms of performance. Then, as $M$ increases, the FD system eventually outperforms the HD system, and both outperform the systems that do not employ an IRS. Similar results can be observed in Fig. \ref{single_UL_DL_rate_vs_SI1}, where with the deployment of the IRS, FD outperforms HD even under strong SI, i.e., $\sigma^2_{\text{SI}}=-42\text{ dBW}$. In addition, the gain over HD system becomes larger with the help of the IRS. To show the superiority of the proposed FD algorithm, we further plot the average DL and UL rate w.r.t. $M$ and $\sigma_{\text{SI}}^2$, respectively, in Fig. \ref{single_UL_DL_rate_vs_SI3} and \ref{single_UL_DL_rate_vs_N2}. These figures indicate that the FD system is able to balance the DL and UL rates. In Fig. \ref{single_UL_DL_rate_vs_SI3}, we can see that, for $-40\text{ dBW}\leq\sigma_{\text{SI}}^2\leq -50 \text{ dBW}$, the reflected link starts to become comparable to the LI link, and thus the SI is effectively mitigated and the UL rate significantly increases.
	
	

	\subsubsection{WSR versus SI power} 
\begin{figure}
		\vspace{-0.5cm}
		\centering
		\includegraphics[width=.5\linewidth]{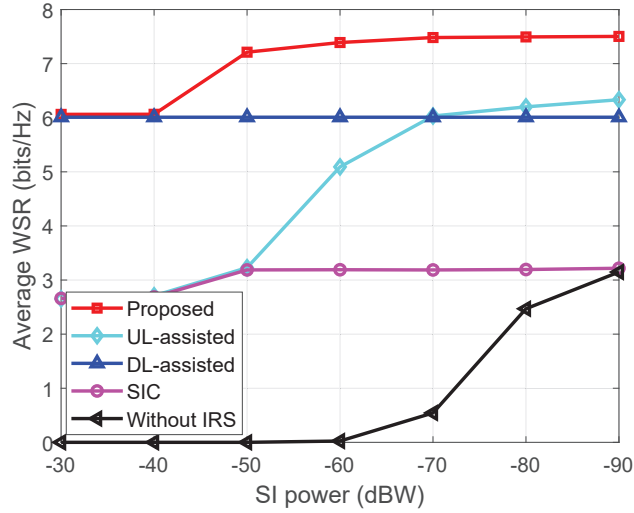}
		\vspace{-0.5cm}
		\caption{Average WSR versus the SI power with $M=200$.}
		\label{single_UL_DL_rate_vs_SI2}
				\vspace{-0.5cm}
	\end{figure}

In Fig. \ref{single_UL_DL_rate_vs_SI2}, we study the average WSR of all methods versus $\sigma_{\text{SI}}^2$ with $M=200$. First, it is observed that, as expected, the WSR of all methods is non-decreasing as SI decreases, and the proposed algorithm enjoys the best performance due to its superiority in interference management. Second, it can be seen that the WSR of the DL-assisted approach remains unchanged, which can be explained as follows. Regardless of the strength of the SI power, with the help of the IRS, the DL rate is generally much larger than the UL rate. Thus again, the DL rate dominates the WSR. Therefore, the overall performance of the DL-assisted approach remains nearly unchanged w.r.t. $\sigma^2_{\text{SI}}$. This again demonstrates the effectiveness of the proposed approach in balancing the DL and UL rates. Third, for the UL-assisted method with $\sigma^2_{\text{SI}}\geq-40 \text{ dBW}$, the WSR is dominated by the DL rate and thus remains unchanged, since the IRS cannot completely suppress the strong SI. When $\sigma^2_{\text{SI}}$ decreases from $-40$ dB to $-70$ dB, the UL rate and WSR increase, whereas with $\sigma^2_{\text{SI}}\leq-70 \text{ dBW}$, the SI becomes weak, so the increase in WSR becomes less noticeable. Fourth, the 		\begin{figure}
		\centering
		\includegraphics[width=.5\linewidth]{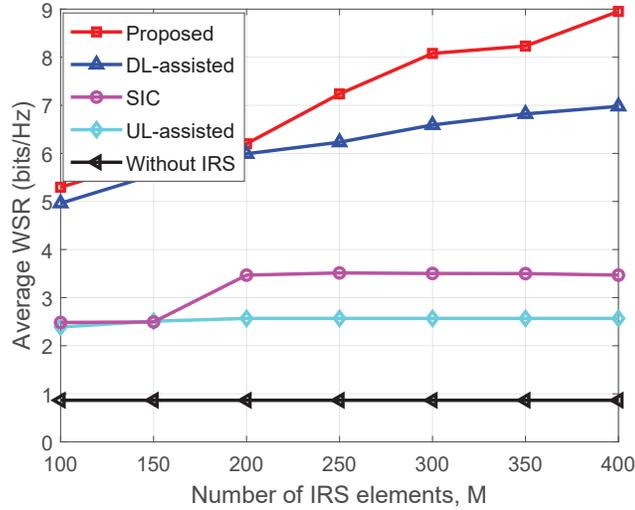}
		\vspace{-0.5cm}
		\caption{Average WSR versus the number of IRS elements with $\sigma_{\text{SI}}^2=-40\text{ dBW}$.}
		\label{single_UL_DL_rate_vs_N3}
		\vspace{-0.5cm}	
	\end{figure}performance of the SIC-based scheme slightly increases, then remains unchanged. Finally, for the approach  without IRS, we see that the increase in WSR comes only when the SI is weak, which is consistent with our intuition. This demonstrates that the IRS is beneficial for SI cancellation. Moreover, it is observed that it approaches the performance of the SIC-based scheme. This is because, when the SI power is low, such SIC-based phase-shift design cannot improve signal transmission, thus the help of IRS becomes negligible. This again shows the superiority of the proposed scheme over other baseline schemes in balancing SI suppression and transmission enhancement.

	\subsubsection{WSR versus number of IRS elements} In Fig. \ref{single_UL_DL_rate_vs_N3}, the average WSR versus $M$ is studied when $\sigma_{\text{SI}}^2=-40 \text{ dBW}$. First, the increase in performance of the proposed approaches increases with $M$ since they balance the need for improving the  transmission efficiency and mitigating the SI. The DL-assisted approach performs similarly to the proposed approach because in this case, the IRS is designed to assist the DL transmission only, leaving the UL transmission unattended, which suffers severely from the strong SI. This is also the reason why the performance of the UL-assisted approach is the worst among all methods that employ an IRS. Finally, we see that the performance of the SIC-based scheme
increases, then saturates, which is explained as in the previous discussion.

	\subsection{General System Setup}
	Next, we consider the general system setup with $N_{\text{T}}=N_{\text{R}}=8$, $K=6$ DL users and $L=2$ UL users. The benchmarks are defined similarly to those in the simple case.
	\subsubsection{WSR versus number of IRS elements}
\begin{figure}
		\vspace{-0.5cm}
		\centering
		\includegraphics[width=.5\linewidth]{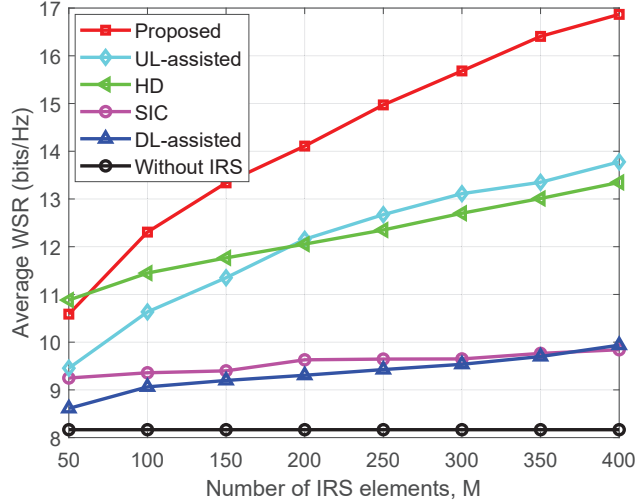}
		\vspace{-0.5cm}
		\caption{Average WSR versus the number of IRS elements with $\sigma_{\text{SI}}^2=-45\text{ dBW}$.}
		\label{multi_rate_vs_N}
		\vspace{-0.5cm}
	\end{figure}

	In Fig. \ref{multi_rate_vs_N}, we compare the average WSR obtained by all methods versus $M$ with			\begin{figure}
	\centering
	\includegraphics[width=.5\linewidth]{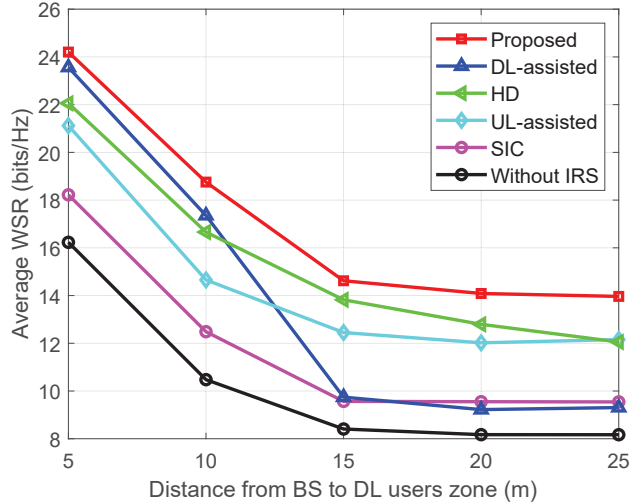}
	\vspace{-0.5cm}
	\caption{Average WSR versus distance from BS to DL users zone with $\sigma_{\text{SI}}^2=-45\text{ dBW}$ and $M=200$.}
	\label{multi_rate_vs_d}
	\vspace{-0.5cm}
\end{figure}
	 $\sigma_{\text{SI}}^2=-45 \text{ dBW}$. The proposed algorithm achieves the highest WSR except when $M=50$, and the performance gain increases with $M$. The HD approach achieves the highest WSR for $M=50$ because the SI and CCI severely degrade FD performance. As $M$ increases, even the UL-assisted approach in the FD system achieves a higher WSR than HD. This implies careful design of the IRS
	 such that it can suppress the SI well and improve transmission quality is crucial in an FD system.

	\subsubsection{WSR versus distance from the BS to the DL users}In Fig. \ref{multi_rate_vs_d}, we study the average WSR versus the distance from the BS to the DL users. The proposed approach again achieves the highest WSR, regardless of the distance.  The WSRs obtained of all approaches decrease as distance increases since the large-scale path loss is greater. The decrease in the WSR of the DL-assisted scheme is the steepest among all methods because, with this design strategy, the IRS is designed only to improve the DL transmission, which is sensitive to the distance from the BS to the DL users. Moreover, it can be seen that when the distance is large, the performance degradation for all approaches is minimal. This is because the UL sum-rate becomes the dominant factor in the WSR in this region, and the decrease in the DL sum-rate becomes negligible.

	\section{Conclusion}
	In this paper, we exploited an IRS to cancel the SI in an FD system. We formulated the optimization problem as one of maximizing the WSR of the DL and UL transmissions under the transmit power constraints. First, to visualize the capability of IRS in SI mitigation by isolating other interference, a simple system setup was considered, where the FD BS serves only one DL and one UL user. To solve this simplified optimization problem, a SCA-based low-complexity algorithm was proposed. Then, we studied the general system model, where the FD BS serves multiple DL and UL users, and an EW-based AO algorithm was proposed for the more challenging optimization problem. Numerical results demonstrated the superiority of the IRS in the proposed FD system. Furthermore, it also 
	showed that the proposed approach is able to balance SI suppression and transmission quality enhancement.

\bibliographystyle{IEEEtran}
\bibliography{FD_IRS}

\begin{thebibliography}{10}
\providecommand{\url}[1]{#1}
\csname url@samestyle\endcsname
\providecommand{\newblock}{\relax}
\providecommand{\bibinfo}[2]{#2}
\providecommand{\BIBentrySTDinterwordspacing}{\spaceskip=0pt\relax}
\providecommand{\BIBentryALTinterwordstretchfactor}{4}
\providecommand{\BIBentryALTinterwordspacing}{\spaceskip=\fontdimen2\font plus
\BIBentryALTinterwordstretchfactor\fontdimen3\font minus
  \fontdimen4\font\relax}
\providecommand{\BIBforeignlanguage}[2]{{%
\expandafter\ifx\csname l@#1\endcsname\relax
\typeout{** WARNING: IEEEtran.bst: No hyphenation pattern has been}%
\typeout{** loaded for the language `#1'. Using the pattern for}%
\typeout{** the default language instead.}%
\else
\language=\csname l@#1\endcsname
\fi
#2}}
\providecommand{\BIBdecl}{\relax}
\BIBdecl

\bibitem{RF1}
M.~Duarte and A.~Sabharwal, ``Full-duplex wireless communications using
  off-the-shelf radios: Feasibility and first results,'' in \emph{Proc.
  Asilomar Conf. Signals, Syst. Comput}, Nov. 2010, pp. 1558--1562.

\bibitem{spectral_efficiency}
D.~Nguyen, L.-N. Tran, P.~Pirinen, and M.~Latva-aho, ``On the spectral
  efficiency of full-duplex small cell wireless systems,'' \emph{IEEE Trans.
  Wireless Commun.}, vol.~13, no.~9, pp. 4896--4910, Jul. 2014.

\bibitem{RF2}
\BIBentryALTinterwordspacing
A.~Sahai, G.~Patel, and A.~Sabharwal, ``Pushing the limits of full-duplex:
  Design and real-time implementation,'' 2011. [Online]. Available:
  \url{https://arxiv.org/abs/1107.0607}
\BIBentrySTDinterwordspacing

\bibitem{RF3}
E.~Everett, M.~Duarte, C.~Dick, and A.~Sabharwal, ``Empowering full-duplex
  wireless communication by exploiting directional diversity,'' in
  \emph{Asilomar Conf. Signals, Systems and Computers}, Nov. 2011, pp.
  2002--2006.

\bibitem{RF4}
T.~Fukui, K.~Komatsu, Y.~Miyaji, and H.~Uehara, ``Analog self-interference
  cancellation using auxiliary transmitter considering {IQ} imbalance and
  amplifier nonlinearity,'' \emph{IEEE Trans. Wireless Commun.}, vol.~19,
  no.~11, pp. 7439--7452, Jul. 2020.

\bibitem{FD1}
J.~I. Choi, M.~Jain, K.~Srinivasan, P.~Levis, and S.~Katti, ``Achieving single
  channel, full duplex wireless communication,'' in \emph{Proc. ACM MobiCom},
  Sep. 2010, pp. 1--12.

\bibitem{FD2}
K.~Singh, K.~Wang, S.~Biswas, Z.~Ding, F.~A. Khan, and T.~Ratnarajah,
  ``Resource optimization in full duplex non-orthogonal multiple access
  systems,'' \emph{IEEE Trans. Wireless Commun.}, vol.~18, no.~9, pp.
  4312--4325, Jun. 2019.

\bibitem{FD3}
U.~Singh, S.~Biswas, K.~Singh, B.~K. Kanaujia, and C.-P. Li, ``Beamforming
  design for in-band full-duplex multi-cell multi-user {MIMO} {LSA} cellular
  networks,'' \emph{IEEE Access}, vol.~8, pp. 222\,355--222\,370, Dec. 2020.

\bibitem{GR}
Q.~Wu and R.~Zhang, ``Intelligent reflecting surface enhanced wireless network
  via joint active and passive beamforming,'' \emph{IEEE Trans. Wireless
  Commun.}, vol.~18, no.~11, pp. 5394--5409, Aug. 2019.

\bibitem{IRS1}
L.~Dai, B.~Wang, M.~Wang, X.~Yang, J.~Tan, S.~Bi, S.~Xu, F.~Yang, Z.~Chen,
  M.~Di~Renzo \emph{et~al.}, ``Reconfigurable intelligent surface-based
  wireless communications: Antenna design, prototyping, and experimental
  results,'' \emph{IEEE access}, vol.~8, pp. 45\,913--45\,923, Mar. 2020.

\bibitem{IRS2}
\BIBentryALTinterwordspacing
J.~Wang, W.~Tang, J.~C. Liang, L.~Zhang, J.~Y. Dai, X.~Li, S.~Jin, Q.~Cheng,
  and T.~J. Cui, ``Reconfigurable intelligent surface: Power consumption
  modeling and practical measurement validation,'' 2022. [Online]. Available:
  \url{https://arxiv.org/abs/2211.00323}
\BIBentrySTDinterwordspacing

\bibitem{coverage_extension1}
M.~Nemati, J.~Park, and J.~Choi, ``{RIS}-assisted coverage enhancement in
  millimeter-wave cellular networks,'' \emph{IEEE Access}, vol.~8, pp.
  188\,171--188\,185, Oct. 2020.

\bibitem{coverage_extension2}
A.-A.~A. Boulogeorgos and A.~Alexiou, ``Ergodic capacity analysis of
  reconfigurable intelligent surface assisted wireless systems,'' in
  \emph{Proc. IEEE 3rd 5G World Forum}, Sep. 2020, pp. 395--400.

\bibitem{coverage_extension3}
G.~Chen, Q.~Wu, C.~He, W.~Chen, J.~Tang, and S.~Jin, ``Active {IRS} aided
  multiple access for energy-constrained {IoT} systems,'' \emph{IEEE Trans.
  Wireless Commun.}, Sep. 2022.

\bibitem{MEC1}
F.~Zhou, C.~You, and R.~Zhang, ``Delay-optimal scheduling for {IRS}-aided
  mobile edge computing,'' \emph{IEEE Wireless Commun. Letters}, vol.~10,
  no.~4, pp. 740--744, Dec. 2020.

\bibitem{MEC2}
T.~Bai, C.~Pan, Y.~Deng, M.~Elkashlan, A.~Nallanathan, and L.~Hanzo, ``Latency
  minimization for intelligent reflecting surface aided mobile edge
  computing,'' \emph{IEEE J. Sel. Areas Commun.}, vol.~38, no.~11, pp.
  2666--2682, Jul. 2020.

\bibitem{MEC3}
Z.~Chu, P.~Xiao, M.~Shojafar, D.~Mi, J.~Mao, and W.~Hao, ``Intelligent
  reflecting surface assisted mobile edge computing for internet of things,''
  \emph{IEEE Wireless Commun. Lett.}, vol.~10, no.~3, pp. 619--623, Nov. 2020.

\bibitem{ISAC1}
M.~Hua, Q.~Wu, C.~He, S.~Ma, and W.~Chen, ``Joint active and passive
  beamforming design for {IRS}-aided radar-communication,'' \emph{IEEE Trans.
  Wireless Commun.}, vol.~22, no.~4, pp. 2278--2294, Apr. 2023.

\bibitem{ISAC2}
Z.-M. Jiang, M.~Rihan, P.~Zhang, L.~Huang, Q.~Deng, J.~Zhang, and E.~M.
  Mohamed, ``Intelligent reflecting surface aided dual-function radar and
  communication system,'' \emph{IEEE Syst. J.}, vol.~16, no.~1, pp. 475--486,
  2021.

\bibitem{ISAC3}
K.~Meng, Q.~Wu, W.~Chen, E.~Paolini, and E.~Matricardi, ``Intelligent surface
  empowered sensing and communication: A novel mutual assistance design,''
  \emph{IEEE Commun. Lett.}, early access, 2023.

\bibitem{multi_cell1}
M.~Hua, Q.~Wu, D.~W.~K. Ng, J.~Zhao, and L.~Yang, ``Intelligent reflecting
  surface-aided joint processing coordinated multipoint transmission,''
  \emph{IEEE Trans. Commun.}, vol.~69, no.~3, pp. 1650--1665, Mar. 2020.

\bibitem{multi_cell2}
C.~Pan, H.~Ren, K.~Wang, W.~Xu, M.~Elkashlan, A.~Nallanathan, and L.~Hanzo,
  ``Multicell {MIMO} communications relying on intelligent reflecting
  surfaces,'' \emph{IEEE Trans. Wireless Commun.}, vol.~19, no.~8, pp.
  5218--5233, May 2020.

\bibitem{multi_cell3}
H.~Xie, J.~Xu, and Y.-F. Liu, ``Max-min fairness in {IRS}-aided multi-cell
  {MISO} systems with joint transmit and reflective beamforming,'' \emph{IEEE
  Trans. Wireless Commun.}, vol.~20, no.~2, pp. 1379--1393, Oct. 2020.

\bibitem{interference_cancellation}
\BIBentryALTinterwordspacing
F.~Wang and A.~L. Swindlehurst, ``Applications of absorptive reconfigurable
  intelligent surfaces in interference mitigation and physical layer
  security,'' 2023. [Online]. Available: \url{https://arxiv.org/abs/2302.01508}
\BIBentrySTDinterwordspacing

\bibitem{FD_SI7}
S.~Tewes, M.~Heinrichs, P.~Staat, R.~Kronberger, and A.~Sezgin, ``Full-duplex
  meets reconfigurable surfaces: {RIS}-assisted {SIC} for full-duplex radios,''
  in \emph{Proc. {IEEE} Int. Conf. on Commun.}, May 2022, pp. 1106--1111.

\bibitem{FD_SI3_omni}
\BIBentryALTinterwordspacing
S.~Fang, G.~Chen, P.~Xiao, K.-K. Wong, and R.~Tafazolli, ``Intelligent omni
  surface-assisted self-interference cancellation for full-duplex {MISO}
  system,'' 2022. [Online]. Available: \url{https://arxiv.org/abs/2208.06457}
\BIBentrySTDinterwordspacing

\bibitem{FD_SI6}
K.~Singh, P.-C. Wang, S.~Biswas, S.~K. Singh, S.~Mumtaz, and C.-P. Li, ``Joint
  active and passive beamforming design for ris-aided ibfd iot communications:
  Qos and power efficiency considerations,'' \emph{IEEE Transactions on
  Consumer Electronics}, vol.~69, no.~2, pp. 170--182, May 2023.

\bibitem{FD_SI2_NF}
\BIBentryALTinterwordspacing
C.~K. Sheemar, S.~Tomasin, D.~Slock, and S.~Chatzinotas, ``Near-field
  intelligent reflecting surfaces for millimeter wave {MIMO} full duplex,''
  2022. [Online]. Available: \url{https://arxiv.org/abs/2211.10700}
\BIBentrySTDinterwordspacing

\bibitem{FD_SI5}
A.~Mondal and S.~Biswas, ``Performance analysis of {RIS} aided {IBFD} {STAR}
  wireless networks,'' in \emph{Proc. IEEE National Conf. on Commun. (NCC)},
  May 2022, pp. 148--153.

\bibitem{FD_SI4_UL}
C.-J. Ku, L.-H. Shen, and K.-T. Feng, ``Reconfigurable intelligent surface
  assisted interference mitigation for {6G} full-duplex {MIMO} communication
  systems,'' in \emph{Proc. IEEE Int. Symp. Pers., Indoor, Mobile Radio Commun.
  (PIMRC)}, 2022, pp. 327--332.

\bibitem{FD_SI1_IB}
W.~Zhang, Z.~Wen, C.~Du, Y.~Jiang, and B.~Zhou, ``{RIS}-assisted
  self-interference mitigation for in-band full-duplex transceivers,''
  \emph{IEEE Trans. Commun.}, early access, 2023.

\bibitem{FD_SI8_connectivity}
R.~Sultan, ``{IoT} connectivity optimization in {RIS}-assisted full-duplex
  massive {MIMO} networks,'' in \emph{Proc. IEEE Symp. on Personal, Indoor and
  Mobile Radio Commun.}, Sep. 2022, pp. 689--694.

\bibitem{CSI}
Y.~Wang, H.~Lu, and H.~Sun, ``Channel estimation in {IRS}-enhanced {mmWave}
  system with super-resolution network,'' \emph{IEEE Commun. Lett.}, vol.~25,
  no.~8, pp. 2599--2603, Aug. 2021.

\bibitem{MMSE_SIC}
D.~Tse and P.~Viswanath, \emph{Fundamentals of Wireless Communication}.\hskip
  1em plus 0.5em minus 0.4em\relax Cambridge University Press, 2005.

\bibitem{socp}
M.~S. Lobo, L.~Vandenberghe, S.~Boyd, and H.~Lebret, ``Applications of
  second-order cone programming,'' \emph{Linear algebra and its applications},
  vol. 284, no. 1-3, pp. 193--228, Nov. 1998.

\bibitem{cvx}
S.~Boyd and L.~Vandenberghe, \emph{Convex optimization}.\hskip 1em plus 0.5em
  minus 0.4em\relax Cambridge University Press, 2004.

\bibitem{convergence}
A.~Beck, A.~Ben-Tal, and L.~Tetruashvili, ``A sequential parametric convex
  approximation method with applications to nonconvex truss topology design
  problems,'' \emph{J. Glob. Optim.}, vol.~47, pp. 29--51, May 2010.

\bibitem{complexity}
K.-Y. Wang, A.~M.-C. So, T.-H. Chang, W.-K. Ma, and C.-Y. Chi, ``Outage
  constrained robust transmit optimization for multiuser {MISO} downlinks:
  Tractable approximations by conic optimization,'' \emph{IEEE Trans. Sig.
  Proc.}, vol.~62, no.~21, pp. 5690--5705, Sep. 2014.

\bibitem{derivative}
J.~Dattorro, \emph{Convex Optimization \& Euclidean distance geometry}.\hskip
  1em plus 0.5em minus 0.4em\relax Palo Alto, CA, USA: Meboo Publishing USA,
  Jun 2010.

\bibitem{rank}
Y.~Huang and D.~P. Palomar, ``Rank-constrained separable semidefinite
  programming with applications to optimal beamforming,'' \emph{IEEE Trans.
  Sig. Proc.}, vol.~58, no.~2, pp. 664--678, Sep. 2010.

\bibitem{zsw}
S.~Zhang and R.~Zhang, ``Capacity characterization for intelligent reflecting
  surface aided {MIMO} communication,'' \emph{IEEE J. Sel. Areas Commun.},
  vol.~38, no.~8, pp. 1823--1838, Jun. 2020.

\bibitem{penalty}
M.~Shao, Q.~Li, W.-K. Ma, and A.~M.-C. So, ``A framework for one-bit and
  constant-envelope precoding over multiuser massive {MISO} channels,''
  \emph{IEEE Trans. Sig. Proc.}, vol.~67, no.~20, pp. 5309--5324, Oct. 2019.

\bibitem{loop_channel}
M.~Duarte, C.~Dick, and A.~Sabharwal, ``Experiment-driven characterization of
  full-duplex wireless systems,'' \emph{IEEE Trans. Wireless Commun.}, vol.~11,
  no.~12, pp. 4296--4307, Nov. 2012.

\bibitem{pathloss}
\BIBentryALTinterwordspacing
M.~Cui, L.~Dai, R.~Schober, and L.~Hanzo, ``Near-field wideband beamforming for
  extremely large antenna array,'' 2021. [Online]. Available:
  \url{https://arxiv.org/abs/2109.10054}
\BIBentrySTDinterwordspacing

\bibitem{near_field}
K.~Dovelos, S.~D. Assimonis, H.~Quoc~Ngo, B.~Bellalta, and M.~Matthaiou,
  ``Intelligent reflecting surfaces at terahertz bands: Channel modeling and
  analysis,'' in \emph{Proc. IEEE Int. Conf. Commun. Workshops (ICC
  Workshops)}, Jun. 2021, pp. 1--6.

\end{thebibliography}
\end{document}